# THE PROPERTIES OF TWO LOW REDSHIFT O VI ABSORBERS AND THEIR ASSOCIATED GALAXIES TOWARD 3C 263[1,2]

B. D. Savage[3], T.-S. Kim[3], B. Keeney[4], A. Narayanan[5],
J. Stocke[4], D. Syphers[4], and B. P. Wakker[3]

ABSTRACT

Ultraviolet observations of the QSO 3C 263 ($z_{em}$ = 0.652) with COS and FUSE reveal O VI absorption systems at z = 0.06342 and 0.14072 . WIYN multi-object spectrograph observations provide information about the galaxies associated with the absorbers. The multi-phase system at z = 0.06342 traces cool photoionized gas and warm collisionally ionized gas associated with a L ~ 0.31L* compact spiral emission line galaxy with an impact parameter of 63 kpc. The cool photoionized gas in the absorber is well modeled with log U ~ -2.6, log N(H) ~17.8, log n(H) ~ -3.3 and [Si/H] = -0.14±0.23. The collisionally ionized gas containing C IV and O VI probably arises in cooling shock heated transition temperature gas with log T ~ 5.5. The absorber is likely tracing circumgalactic gas enriched by gas ejected from the spiral emission line galaxy. The simple system at z = 0.14072 only contains O VI and broad and narrow H I. The O VI with b = 33.4±11.9 km s$^{-1}$ is likely associated with the broad H I λ1215 absorption with b = 86.7±15.4 km s$^{-1}$. The difference in Doppler parameters implies the detection of a very large column of warm gas with log T = 5.61(+0.16, -0.25), log N(H) = 19.54(+0.26, -0.44) and [O/H] = -1.48 (+0.46, -0.26). This absorber is possibly associated with a 1.6L* absorption line galaxy with an impact parameter of 617 kpc although an origin in warm filament gas or in the halo of a fainter galaxy is more likely.
*Short title:* Properties of Two O VI Systems Toward 3C 263

---

[1] Based on observations obtained with the NASA/ESA Hubble Space Telescope, which is operated by the Association of Universities for Research in Astronomy, Inc. under NASA contract NAS5-26555, and the NASA-CNES/ESA Far Ultraviolet Spectroscopic Explorer mission operated by Johns Hopkins University, supported by NASA contract NAS 05-32985.

[2] Based on observations obtained with the WIYN Observatory which is jointly operated by the University of Wisconsin, Indiana University, Yale University and the National Optical Astronomy Observatories.

[3] Department of Astronomy, University of Wisconsin, Madison, WI 53706
[4] CASA, University of Colorado, Boulder, CO
[5] Indian Institute of Space Science and Technology, Thiruvananthapuram 695547, Kerala, India

1. INTRODUCTION

The probing of circumgalactic gas through ultraviolet (UV) absorption lines seen in the spectra of distant QSOs provides the opportunity to study the flow of gaseous matter into and out of galaxies and to search for hidden reservoirs of circumgalactic gas that may play a vital role in galaxy formation and evolution. Such studies are only easily pursued at low redshift where it is possible to clearly identify the galaxy associated with the absorber. However, studies at low redshift require obtaining information on the absorber in the vacuum ultraviolet where the sensitive H I and metal resonance absorption lines occur.

The deployment of the Cosmic Origins Spectrograph (COS) on the Hubble Space Telescope (HST) has greatly expanded the opportunity to study circumgalactic gas because of the factor of 10 to 20 improvement in the throughput of COS compared to earlier HST UV spectrographs (Green et al. 2012; Osterman et al. 2011). The increased throughput allows the study of QSOs well aligned with foreground galaxies of different types. In addition to studying the absorber/galaxy connection, high signal-to-noise (S/N) observations with COS provide important new opportunities to better understand the physical conditions and elemental abundances in the circumgalactic gas at low redshift. Circumgalactic gas can occur in a number of different phases. Cool photoionized gas with log T < 4.7 can be traced by absorption by H I and weakly and moderately ionized metal lines. Warmer collisionally ionized gas can be traced in more highly ionized absorption by C IV, N V, O VI and in some cases Ne VIII. In particularly favorable situations with high S/N observations the warmer gas can also be traced by broad Lyman $\alpha$ (BLA) absorption. When BLA and high ion metal absorption lines can both be seen in an absorption system the difference in the mass between H I and the heavier metals can allow a determination of the relative contributions of thermal and non-thermal absorption line broadening. Measuring the thermal component of the broadening allows a determination of the temperature of the gas and the origin of the ionization in the gas. With sufficient information about the ionization of the warm gas, measures of log N(H I) and log N(O VI) or other high ions can be converted into measures of log N(H) and the metallicity of the gas [O/H] = log[N(O)/N(H)] – log[N(O)/N(H)]$_\odot$. Without the H I reference measurement from the BLA, it would not be possible to determine the metallicity and total baryonic content of the warm gas.

Recent results from COS that illustrate the power of the spectrograph for clearly revealing the existence of warm circumgalactic gas through the presence of BLA and high ionization absorption include those of Narayanan et al. (2010) and Savage et al. (2011a, 2011b). Other studies (Narayanan et al. 2011) have revealed the presence of Ne VIII and O VI in warm circumgalactic gas. The Ne VIII is very difficult to explain by photoionization and therefore requires collisionally ionized gas with log T ~ 5.5 to 6. A very broad BLA and associated narrower O VI absorption has been found in a partial Lyman limit system at z = 0.22601 toward HE 0153-4520 by Savage et al. (2011a). The O VI and BLA imply the direct detection of thermally broadened absorption by hot gas with log T = 6.07 (+0.09, -0.12), [O/H] = -0.28 (+0.09, -0.08) , and log N(H) = 20.41 (+0.13, -0.17). In this case, the galaxy that is likely associated with the absorber has not been identified. However, the general O VI absorber/galaxy association has been clearly established through a number of studies (Stocke et al. 2006; Wakker & Savage 2009; Chen & Mulchaey 2009; Prochaska et al. 2011; Tumlinson et al. 2011b). The various survey projects reveal that galaxy halos containing O VI extend to ~300 kpc with a high covering factor around galaxies with L > 0.1L*. Recent COS studies of the detailed properties of absorption systems containing O VI and information about their associated galaxies include Savage et al. (2010, 2011b), Narayanan et al. (2011), Tumlinson et al. (2011a), and Ribaudo et al. (2011).

In this paper we report on COS and FUSE observations of the z = 0.06342 and 0.14072 O VI absorption systems in the spectrum of the bright QSO 3C 263 with $z_{em}$ = 0.652. The multiphase absorber at z = 0.06342 traces both photoionized and collisionally ionized gas. The absorber at z = 0.14072 mostly traces warm collisionally ionized gas.

In order to identify the galaxies possibly associated with the absorbers, the UV observations are supplemented with observations obtained with the multi-object spectrograph on WIYN of the redshifts of bright galaxies in the 20' radius field surrounding 3C 263. The multi-phase z = 0.06342 absorber is associated with a L ~ 0.31L* compact spiral blue emission line galaxy with an impact parameter of 63 kpc. The z = 0.14072 absorber is possibly associated with a L ~ 1.6L* absorption line galaxy with an impact parameter of 617 kpc although the WIYN measurements at this redshift are only complete to L ~ 0.34L*.

Distances in this paper are physical distances assuming a ΛCDM cosmology with $\Omega_M$ = 0.3, Λ = 0.7 and $H_0$ = 70 km s$^{-1}$ Mpc$^{-1}$ with $h_{70}$ = $H_0$/ (70 km s$^{-1}$ Mpc$^{-1}$)

## 2. UV SPECTROSCOPIC OBSERVATIONS
### 2.1 FUSE Observations

The 3C 263 observations with FUSE (Sahnow et al. 2000; Moos et al. 2003) were obtained through programs G044(53.8 ksec, Shull, PI) and F005 (196.2 ks, Savage, PI). The spectra were processed using the CalFUSE (Ver 2.4) pipeline software. The observations from the different detector segments were combined following the procedures described in Wakker et al. (2003). The zero point of the heliocentric wavelength calibration of the FUSE spectra was established with reference of the ISM UV absorption lines to 21 cm emission in the direction of 3C 263 in the H I survey of Kalberla et al. (2005) with the method described by Wakker et al. (2003). The velocity calibration of the combined FUSE spectrum is estimated to be ±15 km s$^{-1}$. The combined spectrum extends from 912 to 1185 Å but has low S/N for λ <1000 Å. The final combined spectra for λ > 1000 Å only include observations with the LiF channels which have much higher throughput than the SiC channels. The photon count statistical noise per 20 km s$^{-1}$ resolution element in the 1000 to 1150 Å region ranges from 8 to 20. The FUSE observations provide access to the important O VI λλ1031,1037 absorption for the z = 0.06342 system at 1096 and 1102 Å. We assume a Gaussian spectral line spread function with FWHM = 20 km s$^{-1}$ when fitting line profiles to the FUSE observations.

### 2.2 COS Observations

The 3C 263 observations with COS were obtained as part of program GTO/COS 11541 (Green, PI) to study cool, warm, and hot gas in the cosmic web and galaxy halos. Four 3840 s integrations of 3C 263 were obtained with the G130 grating at central wavelengths of 1291, 1300, 1309 and 1318 Å. Four 4500 s integrations were obtained with the G160M grating at central wavelengths of 1589, 1600, 1611, and 1623 Å. The

observations span the wavelength range from 1137 to 1796 Å with a resolution of ~18 km s$^{-1}$. Information about COS can be found in Froning & Green (2009), Green et al (2012) and the COS HST Instrument Handbook (Dixon et al. 2010). The inflight performance of COS is discussed in Osterman et al. (2011) and in numerous instrument science reports found on the Space Telescope Science Institute COS website at http://www.stsci.edu/hst/cos/documents/isrs.

The spectral integrations were obtained with different central wavelengths in order to obtain spectra with different detector/wavelength alignments to reduce the effects of detector fixed pattern noise. The micro-channel plate delay line detector was operated in the time-tag mode with the QSO centered in the 2.5" diameter primary science aperture. The time-tag data were processed with CalCOS version 2.13.6. The full details of how the individual integrations were adjusted for wavelength differences and combined into a single spectrum are discussed by Kim et al. (in preparation). The zero point wavelength offset in the COS observations was established via reference between the COS and FUSE observations of ISM absorption lines and the H I 21 cm emission in the direction of 3C 263. Heliocentric wavelengths and redshifts are reported. The velocity calibration is estimated to be ±10 km s$^{-1}$.

The signal to noise (S/N) in the final combined spectrum ranges from ~20 to 40 per 18 km s$^{-1}$ resolution element. We used the spectral line fitting program VPFIT version 9.5 (http://www.ast.cam.ac.uk/~rfc/vpfit.html), taking into account the wavelength-dependent line spread function of COS. We adopt the Kriss (2011) update to the original parameterization of Ghavamian, P. et al. (2009) which characterizes the core and broad wings of the of the spread function.

## 3. WIYN OBSERVATIONS

The WIYN/HYDRA galaxy redshift survey observations are designed to obtain redshifts for all galaxies with g ≤ 20 that are located within 20' of the 39 QSO sight lines targeted by the COS GTO team and whose photometric redshifts are consistent with being foreground to the QSO. Galaxies are prioritized by their magnitude and proximity to the QSO sight line and, for sight lines like 3C 263 where SDSS spectroscopy exists in the region, highest priority is given to galaxies without SDSS spectra. Galaxies with SDSS redshifts are observed only when no higher-priority targets can be accommodated by a given fiber configuration. Faint (g ≥ 19) galaxies were observed in more than one configuration to improve the S/N.

For the 3C263 sight line, 119 galaxies were found without previously-determined redshifts that met our selection criteria. It was possible to target all but seven of these galaxies by two HYDRA fiber set-up configurations. Each configuration was observed on 8-9 February 2008 in photometric skies with ~0.7" seeing with 3 x 45 min integrations using WIYN/HYDRA with the 600@10.1 grating, the blue cables, and the red bench camera. The data were reduced in IRAF using standard routines and the *dohydra* task for spectral extraction. The resulting spectra have ~4 Å resolution and cover the wavelength range from 3890-6770 Å.

Wavelength calibration was performed using CuAr calibration exposures taken before and after each HYDRA fiber configuration was observed. The calibration exposures were fit by 4th order Chebyshev polynomials and the typical rms of the wavelength solution was ~0.2 Å.

Redshifts were then determined by a custom IDL code and checked by eye for all targets. Of the 120 objects observed, we were able to obtain reliable redshifts for all but three. The median redshift errors are estimated to be ~0.0001 corresponding to ~30 km s$^{-1}$.

Our WIYN/HYDRA survey of galaxies near the 3C 263 sight line is 92% complete (109/119) for galaxies brighter than g = 20 located within 20' of the quasar sight line. All 10 objects for which we were unable to determine redshifts have g $\geq$ 19, so our data are 100% complete to g=19. One of the missed objects has 19 $\leq$ g $\leq$ 19.5 and the other nine have g $\geq$ 19.5. Table1 shows the completeness of our redshift survey as a function of apparent g-band magnitude and angular distance, $\eta$, in arcmin from the quasar sight line. g = [19, 19.5, 20] corresponds to L/L* = [0.15, 0.10 and 0.06] at z = 0.063 and L/L* = [0.85, 0.53 and 0.34] at z = 0.141 using M*(g) = -20.30±0.04 from Montero-Dorta & Prada (2009).

## 4. PROPERTIES OF TWO O VI ABSORPTION SYSTEMS IN THE SPECTRUM OF 3C 263

The COS and FUSE spectra of 3C 263 reveal 5 intervening O VI absorption line systems at z = 0.06342, 0.14072, 0.32568, 0.44674, 0.52683. The system at z = 0.32568 has previously been detected in Ne VIII, O IV, O III using FUSE observations (Narayanan et al. 2010). The full implications of the new COS observations of the multiphase cool and warm gas system at z = 0.32568 revealing H I $\lambda\lambda$1215 to 923, O VI $\lambda\lambda$1031, 1037, C III $\lambda$977, C II $\lambda\lambda$902, 1335, and Si III $\lambda$1207 are presented by Narayanan et al. ( in preparation). In the current paper we discuss the properties of the two lowest redshift O VI systems at z = 0.06342 and 0.14072.

### 4.1. Properties of the z = 0.06342 System

Figures 1a and 1b show absorption line velocity plots for the lines clearly detected in the z = 0.06342 system. The plots display the FUSE and COS observations with ~10 km s$^{-1}$ and ~ 2.3 km s$^{-1}$ binning even though the resolutions are similar (20 km s$^{-1}$ vs 18 km s$^{-1}$). The smaller binning for the COS observations helps in assessing the presence or absence of fixed pattern noise. We adopt the redshift of the O VI $\lambda\lambda$1031, 1037 absorption as the reference redshift of the system. Table 2 lists the absorption line parameters obtained from profile fits using VPFIT. The fitting is performed on the unbinned observations. The multi-phase absorption system is clearly detected in the lines of H I $\lambda\lambda$1215, 1025, 972, O VI $\lambda\lambda$1031, 1037, C IV $\lambda\lambda$ 1548, 1550, C II $\lambda$1334, Si III $\lambda$1206, Si II $\lambda$1260. N V $\lambda$1242 is not detected. There are weak absorbers at 1317.45 and 1317.59 Å that could represent N V $\lambda$1238 absorption. However, the features do not line up with absorption by C IV or O VI. These two features are attributed to H I $\lambda$1215 at z = 0.0837263 and 0.08384. Possible Si IV $\lambda$1402 absorption is severely blended with C III $\lambda$977 in a metal line system at z = 0.52683. Possible Si IV $\lambda$1393 absorption is severely blended with H I $\lambda$1215 at z = 0.21908. C III $\lambda$977 is blended with ISM and terrestrial O I $\lambda$1039 absorption and emission. N III $\lambda$989 is possibly present but the measurement has low significance. N II $\lambda$1083 is blended with ISM P II $\lambda$1152.

Additional information on the line identification and line blending is found in Kim et al. (in preparation).

The well detected lines in the system reveal complex multi-component absorption. The strongest low ionization absorption at v ~ 18-27 km s$^{-1}$ with respect to O VI is clearly revealed in the lines of Si III $\lambda$1206, Si II $\lambda$1260, C II $\lambda$1334. A weaker absorption component is found in Si III $\lambda$1206 at -51 km s$^{-1}$. The C IV $\lambda\lambda$1548, 1550 doublet absorption is well fitted with three blended components at v = 27, -9, and -43 km s$^{-1}$ (see Fig. 1a). The reduced chi square for the fit is $\chi_v^2$ = 1.12. If instead the C IV absorption is fitted with two components, the fit looks poor which is consistent with the larger $\chi_v^2$ = 1.87.

Using the metal line absorption as a guide we have simultaneously modeled the absorption from H I $\lambda\lambda$1215, 1025, 972, and C IV $\lambda\lambda$ 1548, 1550, with three absorption components with the results listed in Table 2 and the fitted profiles shown in Figure 1. In the case of H I, additional components at -161 and +102 km s$^{-1}$ are required to fit the absorption at higher negative and positive velocities. The component at +102 km s$^{-1}$ has been modeled as a single relatively narrow component. However, the positive velocity absorption wing on the H I $\lambda$1215 line for v from 70 to 120 km s$^{-1}$ could instead represent the positive velocity wing of a broad H I Lyman $\alpha$ absorber (BLA) that severely blends with the stronger H I absorption at v < 70 km s$^{-1}$. To test that possibility we modeled the absorption by requiring the existence of an H I absorption component centered at v = 0 km s$^{-1}$, the velocity of the centroid of the O VI absorption. The modeling followed the same procedures as for the profile fit results listed in Table 2. The resulting fit to the H I $\lambda$1215 line is shown in the bottom panel of Figure 1b. The derived fit parameters for the components near -161, -43, -9, and 27 km s$^{-1}$ do not change very much. However, the fit process does yield a BLA with the properties v = 0 km s$^{-1}$, b = 79.2$\pm$15.4 km s$^{-1}$ and log N(H I) = 13.78$\pm$0.20. While the fit to the H I $\lambda$1215 profile near 75 km s$^{-1}$ is better for the solution which includes the single narrow component at +102 km s$^{-1}$ (see Fig. 1b), a solution including the BLA is a reasonable alternate possibility.

The blended high ion component structure is evident in C IV and is not visible in O VI. Therefore, we report results for component fits to the C IV absorption and total column densities for O VI and C IV. The total high ionization absorption column densities are log N(O VI) = 14.59$\pm$0.07 and log N(C IV) = 14.14$\pm$0.05.

The absorption system is clearly revealing multiple gas phases. To determine the properties of the moderately ionized gas in the absorber near 25 km s$^{-1}$, photoionization calculations were performed using Cloudy [ver.C08.00, Ferland et al. 1998] and assuming a uniform density one zone slab model illuminated by the extragalactic background radiation. We use the Haardt & Madau (2001) extragalactic background radiation field incorporating photons from AGNs and star forming galaxies but adjusted to that appropriate for the redshift, z = 0.063 of the absorption system. The ionizing flux log F$_v$ at 1 Rydberg is -21.23, where F$_v$ is in units of erg cm$^{-2}$ s$^{-1}$ Hz$^{-1}$. The hydrogen ionizing photon number density is 1.37 x 10$^{-6}$ cm$^{-3}$.

The assumed model structure is very simple and we apply the model to the strongest absorption component near 25 km s$^{-1}$ detected in H I, Si III, Si II and C II. The heavy element reference abundances assumed in the photoionization model are for the solar photosphere from the abundance compilation of Asplund et al. (2009) and reflect the recent changes in the solar abundances of N, C, and O. Photoionization is a reasonable assumption for the absorber because the H I Doppler parameter of 20.5 km s$^{-1}$ implies an upper limit to the temperature of the gas in the absorber of log T < 4.43 using for the thermal component of the H I line broadening, $b_T = 0.129 (T/A)^{0.5}$ where A is the atomic mass number. A more reliable temperature estimate is obtained by assuming the thermal and non-thermal contributions to the line broadening have Gaussian profiles and add in quadrature with $b^2 = b_T^2 + b_{NT}^2$. We can determine log T and $b_{NT}$ from the observed b values of H I and Si III in the absorber. With b(H I) = 20.5±2.1 and b(Si III) = 12.5±1.2 we obtain logT = 4.22 (+0.16, -0.25) and $b_{NT}$ = 12.1±1.4. At log T = 4.22 the effects of collisional ionization are not important.

The photoionization modeling results are shown in Figure 2 for a model with log N(H I) = 15.15, [O/H] = [N/H] = 0, [C/H] = 0.05, and [Si/H] = -0.14. The different curves show how the expected ionic column densities change with log U where the ionization parameter U = $n_\gamma / n_H$ is the ratio of ionizing photon density to total hydrogen (neutral +ionized) density. The heavy solid lines on the curves for each ion indicate the range of log U where the predicted column density agrees with the observed column density or its limit. The ion column densities in the principal absorption component near v ~ 25 km s$^{-1}$ are log N(H I)= 15.15±0.13, log N(Si III) = 12.90±0.03, log N(Si II) = 12.26±0.09, log N(C II) = 13.37±0.04. The Si III to Si II column densites are most important in determining the properties of the gas.

The observations of H I, Si III, Si II and C II in the principal absorption component are well described by equilbrium photoionization with log U = -2.57±0.10 and [Si/H] = -0.14±0.10±0.13 and [C/H] = +0.05±0.10±0.13. The second error on the abundances mostly comes from the uncertainty in log N(H I). The first error comes from the approximate uncertainty in the application of the simple photoionization model.

The abundances of Si and C in the moderately ionized gas of the absorber are close to solar. These abundances depend on the assumed properties of the background spectrum. If we adopt the pure AGN spectrum of Haardt & Madau (1996) rather than the AGN+ star forming galaxy spectrum, the inferred abundance of Si and C increase by 0.05 and 0.25 dex, respectively.

The foreground galaxy with an impact parameter of 63 kpc and L = 0.31L* ( see §6.1) should not appreciably contribute to the mean intensity of the radiation illuminating the absorber. This follows by scaling the results for the ionizing radiation contributions from the Milky Way for high velocity clouds located at different distances from the Milky Way as presented by Fox et al. (2005) to the galaxy luminosity and absorber/galaxy distance appropriate for the z = 0.063 system. Galaxy G is assumed to have the spectral energy distribution derived for the Milky Way but scaled to the observed luminosity of 0.31L*. The assumptions behind the energy distribution appropriate for the Milky Way are discussed in Fox et al. (2005). We find for Galaxy G, $F_\nu$(galaxy)/$F_\nu$(EGB) ~ 0.16 at 912 Å, implying a small galaxy contribution to the radiation field.

Other derived properties of the moderately ionized gas in the v ~ 25 km s$^{-1}$ absorption component are log T = 3.93±0.05, log L (kpc) = -0.37±0.20±0.13, log N(H) = 17.83±0.10±0.13, log $n_H$ = -3.29±0.10±0.13, log P/k = 0.98±0.10±0.13. When two errors are listed above the first is the approximate modeling error and the second is from the 0.13 dex error in the H I column density. The temperature in the absorber inferred from the photoionization analysis is consistent with the value log T = 4.22(+0.16, -0.25) derived from the observed line widths of H I and Si III.

From Figure 2 we see that photoionized gas with the properties listed above will produce log N(O VI) ~10.6 and log N(C IV) ~ 12.9. For both ions these values are much smaller than the portion of the high ionization absorption that we might reasonably attribute to absorption associated with a component near v ~ 25 km s$^{-1}$. The high ion absorption must arise by a process that is kinematically well connected with the lower ionization absorption. If that process is photoionization, it requires and ionization parameter log U ~ -2.1 to explain the C IV and C II column densities. However, this ionization parameter only yields log N(O VI) ~ 12.3 and predicts 0.25 dex more Si III than is observed. A much larger ionization parameter would be required to produce the O VI. If the O VI and C IV arise in the same gas, log U ~ -0.1 is required to explain the large observed value of log [N(O VI)/N(C IV)] ~0.45 for the entire C IV and O VI absorption. This ionization parameter is 2.6 dex larger than that required to explain the cool photoionized gas containing Si II, Si III, and C II. Such a large ionization parameter would require a 2.6 dex density decrease between the cool photoionized gas and the much lower density gas producing C IV and O VI. Such a situation seems unlikely given the strong kinematical connection between the low and high ionization absorbers.

The strong kinematic connection between the low and high ionization absorption suggests to us that the high ionization absorption occurs in some type of cool/hot gas interface between the low ionization (cool gas) and a possible hot exterior medium. Given the kinematical complexity of the absorption we unfortunately do not have reliable information about possible broad Lyman $\alpha$ absorption that might trace the small amount of thermally broadened H I that could be associated with the O VI and C IV absorption. However, if the BLA with b(BLA) = 79.2±15.4 km s$^{-1}$ shown in the bottom panel of Figure 1b actually exists, it could trace the small amount of H I existing in the warm gas responsible for the O VI absorption with b(O VI) = 38.7±6.3 km s$^{-1}$. The difference in the H I and O VI line widths imply log T = 5.49( +0.19, -0.27) and $b_{NT}$ = 34.3 (+2.1, -2.7) km s$^{-1}$ if the H I and O VI reside in the same uniform temperature plasma. Log T = 5.49 is near the temperature where O VI peaks in abundance under the conditions of collisional ionization equilibrium (CIE). If CIE is a good description of the ionization we can obtain log N(O) and log N(H) from log N(O VI) = 14.59±0.07 and log N(H I) = 13.78±0.20 by using the ionization corrections from Gant & Sternberg (2007). We obtain log N(O) = 15.28 (+2.0,+0.07, -0.00, -0.07) and log N(H) = 19.65 (+0.32, +0.20, -0.54, -0.20) where the first error is from error in the ionization correction due to the temperature uncertainty and the second error is from the H I or O VI column density error. Adding the errors in quadrature we obtain log N(O) = 15.28 (+2.0, -0.07) and log N(H) = 19.65 (+0.38, -0.58). The value for the total oxygen column density has a very large positive error because the abundance of O VI in CIE decreases very rapidly for log T < 5.5. Therefore, the abundance of oxygen in the warm plasma is poorly constrained to [O/H] = -1.06 (+2.56, -0.21) using log(O/H)$_\odot$ = -3.31 from Asplund et al. (2009). However, the total hydrogen column density log N(H) = 19.65(+0.38, -0.58) is relatively well determined. If the parameters for the BLA are correct, the column of gas in the BLA is ~66 times larger than the column in the cool photoionized gas with log N(H) = 17.83±0.16.

### 4.2. Properties of the z = 0.14072 System

Figure 3 shows absorption line velocity plots for the lines clearly detected in the z = 0.14072 system. Figure 4 shows flux versus wavelength for the region from 1382 to 1392 Å centered on the H I λ1216 line near 1386.8 Å. The H I profile fit window and continuum placement windows are illustrated. A low significance weak feature near 1387.6 Å is probably not real. H I λ972 absorption in a system at z = 0.42690 containing H I λλ1215, 1025 is expected near 1377.71Å but with a strength much less than the low significance feature. The very weak depression shown in the dashed continuum displayed in Figure 4 near 1377.7 Å shows the expected line strength for the z = 0.42690 H I λ972 line.

The continuum for the H I λ1216 line near 1386.8 Å is well defined and represented by a first order Legendre polynominal. The fit parameters for a single component fit to O VI and a two component fit to H I are given in Table 3 with the fit curves illustrated in Figure 3.

The highly ionized absorption system only contains O VI λλ1031, 1037 and H I λλ1215, 1025. Lines not detected include: Si II λ1260, Si IV λλ1394, 1403, C II λ1334, C III λ977, C IV λ1550, and N V λλ1238, 1242. C IV λ 1548 is severely blended with H I λ 1215 at z = 0.45279. An absorption line that could be identified as Si III λ1206 at z = 0.1407 appears near 1376.3 Å. However, it is unlikely to be Si III because C III λ977 which would be expected to be much stronger is not detected in the FUSE spectrum at 977.0Å. We identify the line at 1376.3Å as Ly λ1215 at z = 0.13209.

The system is kinematically very simple. The H I absorption reveals a narrow and broad component with b(H I) = 27.9±1.0 and 86.7±14.6±4.8 km s$^{-1}$ and log N(H I) = 14.51±0.03 and 13.47±0.10, respectively. The two component H I fit is required to explain the broad wings on the H I λ1215 absorption profile. $\chi_\nu^2$ is 2.44 for the single component fit and 1.72 for the two component fit. The 0.10 dex column density error on the broad component implies a detection significance of ~3.8 sigma. The uncertainty for the width of the BLA is dominated by the profile fit error of 14.6 km s$^{-1}$. The continuum placement uncertainty for the fit is estimated to be an additional ± 4.8 km s$^{-1}$ by selecting many different >0.4 Å wide continuum regions in the continuum windows shown in Figure 4 and deriving the BLA fit parameters for each selected continuum region. When deriving physical properties of the plasma we assume the profile fit error and continuum error on the Doppler width of the BLA add in quadrature and adopt b(BLA) = 86.7± 15.4 km s$^{-1}$.

The narrow and broad H I absorption components differ in velocity by 14±7 km s$^{-1}$, while, v(O VI) –v(BLA) = 7 ± 17 km s$^{-1}$ and v(O VI) – v(narrow H I) = -7±16 km s$^{-1}$. Within the fit and velocity calibration errors the O VI absorption could be associated with either the narrow H I absorption or the BLA. The O VI absorption is fitted with b(O VI) = 33.4±11.9 and log N(O VI)= 13.60±0.09.

The difference in the H I and O VI line widths can be used to determine the temperature of the absorbing plasma using for the thermal component of the line broadening, $b_T = 0.129\,(T/A)^{0.5}$ where A is the atomic mass number. Assuming the thermal and non-thermal contributions to the broadening have Gaussian profiles and add in quadrature with $b^2 = b_T^2 + b_{NT}^2$, we can determine log T and $b_{NT}$ from the b values for O VI and H I if these two species arise in the same gas assumed to be at a uniform temperature. The O VI absorption must have $b(O\ VI) < b(H\ I) = 27.9\pm1.0$ km s$^{-1}$ for it to be associated with the narrow H I absorption. Unfortunately the Doppler parameter for O VI is uncertain with $b(O\ VI) = 33.4\pm11.9$ km s$^{-1}$. If we take the true value of the O VI Doppler parameter to be $33.4-11.9 = 21.5$ km s$^{-1}$ we obtain log T = 4.31 and $b_{NT}$ = 21 km s$^{-1}$ for the plasma containing O VI if it is associated with the narrow H I absorption. In this case, the gas temperature is so low that the O VI would arise by photoionization from the extragalactic background radiation in a very low density medium. Given the uncertainty of the O VI Doppler parameter we can not rule out this possibility. However, such an interpretation leaves the origin of the BLA unexplained. It, therefore, appears more likely that the O VI is directly associated with the BLA rather than with the narrow H I absorption.

Assuming the O VI and BLA exist in the same uniform temperature gas we obtain from $b(H\ I) = 86.7\pm15.4$ km s$^{-1}$ and $b(O\ VI) = 33.4\pm11.9$ km s$^{-1}$, log T = 5.61(+0.16, -0.25) and $b_{NT}$ = 26.3 $\pm16.5$ km s$^{-1}$. The plasma traced by the O VI and the broad H I absorption is warm.

Knowing the plasma temperature we can convert the values of log N(H I) and log N(O VI) into log N(O) and log N(H) and determine the oxygen abundance and total baryonic content of the gas. We assume collisional ionization equilibrum which is a very good approximation at the large implied temperature and low implied abundance of oxygen (Gant & Sternberg 2007). From log N(H I) = 13.47±0.10, log N(O VI) = 13.60±0.09 and log T = 5.61(+0.16, -0.25) we obtain log N(H) = 19.54(+0.10+0.24, -0.10-0.43) and [O/H] = -1.48 (+0.13+0.44, -0.13-0.23). In these results the first error reflects the uncertainty in the column densities while the second error reflects the uncertainty in the ionization correction caused by the uncertainty in the estimated temperature. Adding the errors in quadrature we obtain log N(H) = 19.54(+0.26, -0.44) and [O/H]=-1.48 (+0.46, -0.26). The warm gas in this system is very highly ionized with log[N(H)/N(H I)] = 6.10 (+0.35, -0.50). The absorber is tracing gas with a very large content of baryons and a relatively low metallicity.

## 5. GALAXIES ASSOCIATED WITH THE TWO ABSORBERS

Figure 5 displays an 40' x 40' SDSS r-band mosaic of the region surrounding the QSO 3C 263 with the positions of the galaxies within 500 km s$^{-1}$ of the two absorbers marked along with the redshifts measured by WIYN. The star marks the position of the QSO. The two circles centered on the QSO have radii of 200 kpc at z = 0.063 (large circle) and at z = 0.140 (small circle). The properties of the galaxies found near z = 0.063 and z = 0.140 are listed in Tables 4 and 5.

### 5.1. Galaxies Associated with the z = 0.06342 System

The WIYN observations (see Figure 5 and Table 4) reveal 5 galaxies near z ~ 0.063 with impact parameters ranging from 63 to 1210 kpc and $\Delta v < 500$ km s$^{-1}$. The galaxy closest to the line of sight to the QSO has an impact parameter of 63 kpc and a redshift of 0.06322±0.00013. The centroid of the O VI absorption is at z = 0.06342 which differs from the galaxy redshift by $\Delta z = 0.00020\pm0.00013$ which corresponds to $\Delta v = -56\pm39$ km s$^{-1}$ in the rest frame of the absorption system. This galaxy (hereafter Galaxy G) is most likely associated with the absorber. Galaxy G lies in a group of 5 galaxies with L

= 0.28L* to 0.07 L* extending over ~2 Mpc (see Figure 5 and Table 4) . The average velocity of the group with respect to the absorber is -43 km s$^{-1}$ and the velocity dispersion of the group is 250 km s$^{-1}$.

Figure 6 shows a 2000 second high resolution HST/WFPC2 image of the 3C 263 field in the F675W filter (Proposal ID: 5978, PI: S. Rawlings), reduced with the MULTDRIZZLE package in IRAF. 3C 263 and Galaxy G are identified. The insert in the upper right side of the figure shows an expanded image of Galaxy G. Galaxy G is a compact (optical extent ~ 4 kpc) nearly face-on spiral galaxy with a bright nuclear region and several bright knots along its spiral arms.

A second galaxy evident in the HST image lies on top the radio loud QSO 3C 263. As the QSO is very bright, the details of this galaxy's morphology should be addressed after a deconvolution of the HST point spread function. However, the galaxy is bright enough it can be recognized through a careful inspection of the existing observations to contain a faint asymmetric tail-like feature in the south east and an extension to the north. Many host galaxies of radio-loud QSOs show a recent merger or interaction indicated by tidal tails and distorted morphology (Benner et al. 2008; Cales et al. 2011). Therefore the galaxy overlying the position of 3C 263 is probably the QSO host galaxy.

Galaxy G has SDSS model Mag values of u = 19.13±0.03, g = 18.24±0.01, r = 17.89±0.01, i = 17.56±0.01, and z = 17.48±0.02. With M(x) = x-37.26 for z = 0.063 we determine absolute magnitudes, M(u) =-18.13, M(g) = -19.02, M(r) = -19.37, M(i) = -19.70, and M(z) = -19.78. Using M*(g) = -20.30±0.04 from Montero-Dorta & Prada (2009) we find L = 0.31L* for Galaxy G.

The WIYN/HYDRA spectrum of Galaxy G is shown in Figure 7 after corrections for the foreground Galactic extinction with E(B-V) = 0.011 mag and A$_V$ =0.037 mag (Schlegel, Finkbeiner & Davis 1998) . The galaxy is blue with relatively strong emission lines of Hβ, Hγ, Hδ, He I λ5876, [O I] λ6300, [O II] λ3727, and [O III] λλ4959, 5007. Unfortunately several important diagnostic lines are not covered in the spectrum including Hα and the density sensitive [S II] λλ6716, 6731 and temperature sensitive [N II] λλ5755, 6548, 6583 lines. The density sensitive [O II] λλ3726, 3729 emission is covered but with inadequate resolution. Absorption lines from Hβ, Hγ, Hδ, Hε, and higher Balmer series lines and from Ca II λλ3933, 3968 are also present. The integrated spectrum reveals a typical, low-excitation oxygen-rich star-forming galaxy with active star formation within several Gyr. The non-detection of [O III] λ4363 implies a relatively high metallicity [O/H] > -0.3 (Kewley & Dopita 2002).

There is a strong underlying integrated stellar spectrum which is dominated by A stars indicating strong star formation over several Gyrs. We used the analysis code kindly provided by Christy Tremonti (see Tremonti et al. 2004 for details) to model the stellar continuum. The code estimates the continuum via a linear combination of different model stellar spectral energy distributions. Since the galaxy has a solar metallicity (see below) we used the simple Stellar Population models of the GALAXEV 2003 library of Bruzual & Charlot (2003) with solar metallicity and 10 different ages. We subtracted the modeled stellar continuum from the observed one in order to correct for stellar absorption in the emission line intensities for the Balmer lines.

Since the WIYN/MOS spectrum of Galaxy G is the integrated spectrum of an inhomogeneous ensemble of different H II regions, it is not straightforward to derive an accurate abundance for the galaxy, nor the dust extinction. We follow the conventional procedure adopted by Tremonti et al. (2004) to derive both parameters.

The internal dust extinction is corrected in the rest frame, assuming the galaxy-averaged attenuation law $\tau_\lambda = a\ \lambda^{-0.7}$ (Charlot & Fall 2000; Dopita et al. 2006). The value of the constant a is determined from the observed Hγ/Hβ emission line ratio since Hα was not covered by the spectrum. Case B recombination is assumed along with an electron density and temperature of $10^4$ cm$^{-3}$ and $10^4$ K (Osterbrock 1989). The observed reddened Hγ/Hβ emission ratio is 0.447±0.020 where the error reflects the continuum placement uncertainty. This implies a visual optical depth of $\tau_V = 0.59\pm0.04$ or a visual extinction of $A_V = 0.54\pm0.04$. The observed and de-reddened emission line intensities are listed in Table 6.

We measured the $R_{23}$ value, $R_{23} = (I_{[O\ III]\lambda3727} + I_{[O\ III]\lambda\lambda4959,5007})/I_{H\beta}$, and employed the $R_{23}$-abundance relation of Tremonti et al. (2004),

$12+\log(O/H) = 9.185 - 0.313x - 0.264x^2 - 0.321x^3$, where $x = \log R_{23}$.

This abundance calibrator is valid for the upper branch of the double-valued $R_{23}$-abundance relation for galaxies without detected [O III] λ4363 emission as is the case for Galaxy G. With the observed de-reddened intensities listed in Table 6, we obtain the oxygen abundance $12+\log(O/H) = 8.8\pm0.1$. The error is estimated from the width of the oxygen abundance distribution at a given $R_{23}$ in Figure 3 of Tremonti et al. (2004) delineated by the Zaritsky, Kennicutt & Huchra (1994) and the McGaugh(1991) $R_{23}$-abundance relations. With the uncertain extinction correction and the mixing of the physical properties of spatially different individual H II regions, the uncertainty in the oxygen abundance could be as large as 0.2 dex. Taking the solar oxygen abundance as $12+\log(O/H)_\odot = 8.69$ (Asplund et al. 2009) we obtain [O/H] = 0.11±0.20 dex for the abundance of oxygen in Galaxy G.

The star formation rate of Galaxy G can be estimated from the reddening-corrected intensity of Hβ. For Case B recombination at $T = 10^4$ K and $n_e = 10^4$ cm$^{-3}$, $I_{H\alpha}/I_{H\beta} = 2.85$. With $I_{H\beta} = 148.6$ erg cm$^{-2}$s$^{-1}$ we estimate $I_{H\alpha} = 424$ erg cm$^{-2}$ s$^{-1}$ and a Hα luminosity $L(H\alpha) = 4.2\times10^{41}$ erg s$^{-1}$. Using the star formation rate (SFR) and $L(H\alpha)$ conversion of Kennicutt (1998), we obtain the star formation rate of Galaxy G as SFR ($M_\odot$ yr$^{-1}$) = $7.9\times10^{-42}$ L(Hα) = 3.2 $M_\odot$ yr$^{-1}$. This value should be considered a lower limit since the HYDRA 3" fiber doesn't capture light from the full extent of the galaxy.

Galaxy G is estimated to have a stellar mass of $M^* = (4.6\pm1)\times10^9\ M_\odot$ kindly calculated by Dr. Yamnei Chen using the techniques described in Chen et al. (2011).

In summary: Galaxy G is a 0.31L* post starburst spiral galaxy with a current star formation rate of 3.2 $M_\odot$ yr$^{-1}$. Its stellar mass is $M^* = (4.6\pm1)\times10^9\ M_\odot$ implying a total (star+ halo) mass of $\sim10^{11}\ M_\odot$. With an oxygen abundance [O/H] = 0.11±0.20 and a stellar continuum dominated by Balmer line absorption from A stars, it is reasonable to assume the multi-phase absorber is tracing halo gas strongly affected by matter ejected from the galaxy over the past several Gyrs.

## 5.2. Galaxies Associated with the z = 0.14072 System

The WIYN redshift survey has revealed 6 galaxies with redshifts ranging from 0.13998 to 0.14194 in the 40'x40' field centered on 3C 263 (see Figure 5 and Table 5). The two closest galaxies have impact parameters of 617 kpc and 1.27 Mpc. The other 4 galaxies have impact parameters ranging from 1.6 to 3.4 Mpc. The absorber may be associated with the galaxy with an impact parameter of 617 kpc which has a redshift of 0.14087±0.00023. The galaxy to absorber redshift difference implies a small velocity difference of 38±71 km s$^{-1}$ in the rest frame of the system. The galaxy with the impact parameter of 617 kpc lies in an group of 6 galaxies with an average galaxy to absorber velocity of 97 km s$^{-1}$ and a velocity dispersion of 169 km s$^{-1}$. The group extends over a region ~ 5 Mpc in diamater (see Figure 5).

The galaxy at z = 0.14087 has SDSS model Mag values of u = 20.40±0.10, g = 18.30±0.01, r = 17.25±0.01, i = 16.82±0.01, and z = 16.48±0.01. With M(x) = x-39.12 for z = 0.141 we determine absolute magnitudes, M(u) =-18.72, M(g) = -20.82, M(r) = -21.87, M(i) = -22.30, and M(z) = -22.64. Using M*(g) = -20.30±0.04 from Montero-Dorta & Prada (2009) for H$_o$ = 70 km s$^{-1}$ Mpc$^{-1}$ we find L = 1.61 L* for the galaxy. The SSDS photometric redshift of the galaxy, z = 0.132±0.010, agrees well with the measured redshift.

The WIYN spectrum of the galaxy at z = 0.14087 is shown in Figure 8. The galaxy has a pure absorption line spectrum with the redshifted lines of Hβ, Hγ, Ca II λλ3933, 3968, G band, Mg I λλ5173, 5184, and Na I λλ5890, 5896 marked on the figure. There is no evidence of [O II] λ3727 or [O III] λ5007 emission. Possible Hα λ6563 emission is redshifted beyond the long wavelength coverage of the observation. With reference to the Kennicutt (1992) observations of the integrated spectra of nearby galaxies, the observed spectrum is broadly consistent with a classification ranging from E4 to Sa with no evidence for substantial amounts of current star formation.

The WIYN redshift survey is 98% complete to g = 20.0 within 10' of 3C 263 (see Table 1). The magnitude g = 20.0 corresponds to a L = 0.34L* galaxy at the redshift of 0.141. It is therefore possible that the absorber is not related to the galaxy with the impact parameter of 617 kpc but is instead associated with a unseen fainter galaxy having L < 0.34L*. A much deeper redshift survey to g ~ 22 will be required to rule out the absence of galaxies associated with the absorber as faint as L = 0.05L*.

## 6. ORIGINS OF THE ABSORBERS
### 6.1. The z = 0.06342 System

The absorber very likely arises in enriched circumgalactic gas surrounding Galaxy G with L ~ 0.27L$_*$ and an impact parameter of 63 kpc. The cool photoionized gas in the absorber and the star forming regions in the galaxy share similar near solar abundances. The spectrum of the galaxy is dominated by A stars implying the galaxy underwent an intense phase of star formation several Gyrs ago. In addition, the current star formation rate of 3.2 M$_\odot$ yr$^{-1}$ and the stellar mass of 4.6x10$^9$M$_\odot$ imply a high current specific star formation rate of 7x10$^{-10}$ yr$^{-1}$.

The highly ionized gas in the absorber may represent cooling transition temperature gas that was heated during an earlier starburst wind driven ejection process. An alternate possibility is the warm transition temperature gas arises in the interface/interaction zone between the cool photoionized gas of the absorber and a hot (unseen) exterior medium. The very good kinematical relationship between the warm gas tracers (C IV and O VI) and the cool gas tracers (Si II, Si III and C II) lends support to the origin of the O VI and C IV in interface/interaction zones.

The Wakker & Savage (2009) study of very low redshift O VI absorption with cz < 5000 km s$^{-1}$ showed that O VI absorption is detected in 70% of L > 0.1L* field galaxies with impact parameters less than 350 kpc.   The recent Tumlinson et al. (2011b) COS survey of O VI absorption associated with star forming galaxies with $\log(M/M_\odot)$ ranging from 9.6 to 11.35 revealed the ubiquitous presence of O VI halos with $\log N(O\ VI) > 14.5$ in the entire sample of 42 galaxies with impact parameters <150 kpc. Although the stellar mass of Galaxy G with $\log(M/M_\odot) = 9.6$ lies at the low end of the Tumlinson et al. sample, its specific star formation rate of $7 \times 10^{-10}$ yr$^{-1}$ is higher than for any galaxy in the sample.  It is therefore quite likely Galaxy G has an extensive O VI halo with a large covering factor.

The cool photoionized gas in the absorber at z = 0.063 has $\log N(H) \sim 17.8$ and [Si/H] $\sim -0.14$.  The warm collisionally ionized gas has $\log N(O\ VI) = 14.59 \pm 0.09$.  If the oxygen abundance in the warm gas is close to solar as in the cool gas, we can estimate the total hydrogen column density in the absorber.  With $\log(O/H) = \log(O/H)_\odot = -3.31$ (Asplund et al. 2009), $\log N(H) = \log N(O\ VI) + 3.31 + \log(O/O\ VI)$, where (O/O VI) is the ionization correction.  O VI is relatively easily ionized.  Therefore, the value of $\log(O/O\ VI)$ in collisionally ionized gas is >0.7 (Gnat & Sternberg 2007).  Therefore $\log N(H) > 18.6$ in the warm ionized circumgalactic gas along the path to 3C 263 at z = 0.06342.  The column density of warm gas exceeds that of the cool gas by more than a factor of ~6.  If the possible BLA discussed in section 4.1 is real and traces the plasma producing the O VI absorption the temperature of the gas is estimated to be $\log T = 5.49(+0.19, -0.27)$.  This implies a total column density of hydrogen of $\log N(H) = 19.65 (+0.38, -0.58)$ and a baryonic content of the warm gas exceeding that in the cool gas by a factor of ~66.

If the O VI halo surrounding Galaxy G extends to radius R, the total mass of warm ionized gas in the halo assuming a solar oxygen abundance is approximately given by $M_{gas} = 4\pi R^2 N(H) 1.33\ m_H (O/O_\odot)$, where $m_H$ is the mass of hydrogen and the 1.33 corrects for the presence of He.   With $\log N(H) > 18.6$ and R = 100 kpc, we obtain a halo warm gas mass of $M_{gas}/M_\odot \geq 5.5 \times 10^9 (R/100\ kpc)^2 (O/O_\odot)$.   This simple estimate reveals the mass of warm gas mass in the halo out to ~100 kpc is likely larger than the total mass of stars in the galaxy, estimated to be $4.6 \times 10^9 M_\odot$.  The gas in the halo of Galaxy G will likely play a major role in the future evolution of the galaxy.

The measurements provide direct evidence that modest luminosity galaxies with evidence for vigorous star formation over several Gyrs have strong enough mass ejection mechanisms to pollute their circumgalactic surroundings to large distances.

### 6.2. The z = 0.14072 System

The WIYN galaxy redshift survey has revealed the galaxy with the smallest impact parameter from 3C 263 near the redshift of the absorber is a luminous absorption line galaxy with $\rho = 617$ kpc, L = 1.61L* and no evidence for current star formation.   The absorber and galaxy have nearly identical redshifts implying a restframe absorber to galaxy velocity difference of only $38 \pm 71$ km s$^{-1}$.  The multiphase absorber contains cool and warm H I.  The warm H I and O VI traces a plasma with $\log T = 5.61\ (+0.16, -0.25)$, $\log N(H) = 19.54\ (+0.26, -0.44)$ and $[O/H] = -1.48\ (+0.46, -0.26)$. If the absorber is associated with the absorption line galaxy, it could trace a warm highly extended halo or circumgalactic gas that was shock heated to its current temperature during the galaxy formation process.

However, it is possible the cool/warm gas absorber at z = 0.14072 is not associated with the absorption line galaxy with $\rho$ = 617 kpc. The WIYN survey is complete to the 100% level at g = 20 within 10' of 3C 263 at the redshift of the absorber. However, g = 20 corresponds to L = 0.34L* at z = 0.141. Therefore, the absorber could instead be associated with a unseen fainter galaxy having L < 0.34L*.

If additional fainter galaxies are not found near z = 0.141, the absorber could represent the detection of the warm plasma in a WHIM filament. The pie chart in Figure 9 shows a 2-degree wedge in RA and Dec centered on the 3C 263 sight line. The various letters mark the positions of galaxies from the CfA redshift survey (c), the SDSS survey (s) and from the WIYN survey of this paper (k). Magenta symbols identify galaxies with impact parameters of less than 1 Mpc from the line of sight. The z=0.14072 absorber denoted with the circle at cz = 44,100 km s$^{-1}$ does not appear to originate in a particularly overdense environment such as that seen at z = 0.063 corresponding to cz = 18,900. However, the observations from the three redshift surveys shown in Figure 9 suffer from varying degrees of incompleteness for cz > 35,000 km s$^{-1}$ and it is possible the wedge plot may only be revealing the brighter members of the galaxies in a filament near cz = 44,000 km s$^{-1}$. The presence of the 6 galaxies listed in Table 5 with impact parameters ranging from 617 to 3.4 Mpc and luminosities from 0.29L* to 1.6L* reveals the region does contain a number of moderate luminosity galaxies.

The properties of the absorber at z = 0.14072 (narrow and broad H I and O VI) are similar to those for the absorber at z = 0.01028 observed toward Mrk 290 by Narayanan et al. (2010). That absorber also is only detected in H I and O VI. The H I absorption reveals two components separated by 114± 5 km s$^{-1}$ with b = 55±1 and b = 33±1 km s$^{-1}$. The broad H I absorption with log N(H I) = 14.35±0.01 is well aligned with the O VI absorption having b(O VI) = 29±3 km s$^{-1}$ and log N(O VI) = 13.80±0.05. The detailed analysis of the Mrk 290 BLA and O VI absorption implies the existence of a highly ionized plasma with log T = 5.15, [O/H] ~ -1.7, log N(H) ~ 19.6. The gas is not in collisional ionization equilibrium. Therefore, the values of the derived parameters are more uncertain than for the z = 0.1407 system toward 3C 263. However, the inferred low oxygen abundance and high total hydrogen column density are comparable for the Mrk 290 and 3C 263 absorbers. The line of sight to the Mrk 290 absorber is within the SDSS footprint with galaxy redshifts complete to g = 17.5 which corresponds to L~ 0.03L* at z = 0.01. The Mrk 290 absorber passes through a galaxy filament with more than two dozen galaxies within an impact parameter of 1.5 Mpc (see Figure 4 in Narayanan et al. 2010). The nearest galaxy to the line of sight with an impact parameter of 424 kpc is NGC 5987, a 2L* galaxy with Sb morphology. NGC 5987 is one of the largest spiral galaxies in the nearby universe. Since no faint galaxies with L > 0.03L* are detected near the absorber, the BLA+O VI system toward Mrk 290 is either tracing the very distant outer regions of NGC 5987 or warm filament gas that may connect to NGC 5987. A similar filament/galaxy connection may exist for the z = 0.14072 absorber toward 3C 263 although in this case it is less likely that the absorber is related to the absorption line galaxy at 617 kpc. Systematically determining the total extents of O VI halos surrounding starforming and absorption line galaxies will be important for discriminating between the different possible origins.

# 7. DISCUSSION

To ultimately understand the origin of absorption systems containing O VI and to estimate their baryonic content it is extremely important to measure elemental abundances in the systems. This can only be done if there is a measure of the column density of H I in the same gas producing the O VI absorption. If the O VI exists in warm gas with log T = 5 to 6, the associated H I absorption will be relatively broad, relatively weak and difficult to detect. Table 7 lists the O VI systems so far studied that trace warm gas for which measures of the associated BLA have been obtained. These are the only warm gas systems for which we currently have any information about the elemental abundances and the total baryonic content in the warm gas near galaxies.

All the systems exhibit very large total H column densities with log N(H) ranging from 19.6 to 20.4. The oxygen abundance ranges from [O/H] = -0.28 to -1.7. Definite information on the associated galaxy only exists for the system at z = 0.20701 toward HE 0226-4110 (Mulchaey & Chen 2009; Savage et al. 2011b) . For HE 0153-4520 the follow up galaxy redshift measurements with WIYN have been obtained but the observations do not go deep enough to be useful for identifying galaxies with L < 0.3L* at z = 0.226. For the system toward MRK 290 the galaxy with the smallest impact parameters is at 424 kpc. In this case, the absorber may be tracing the gas in intergalactic filaments (Narayanan et al. 2010).

# 8. SUMMARY

Ultraviolet observations of 3C 263 ($z_{em}$ = 0.652) with COS and FUSE reveal multiphase O VI absorption systems at z = 0.06342 and 0.14072. The observations are used to determine the physical properties of the gas in the systems. WIYN multi-object spectrograph observations provide information about the galaxies associated with the absorbers.

1. The multi-phase system at z = 0.06342 containing H I, C II, Si II, Si III, C IV, and O VI traces cool photoionized gas and warm collisionally ionized gas . The photoionized gas in the absorber is well modeled with log U ~ -2.6 , log N(H) ~ 17.8, log L(kpc) ~ -0.37, log n(H) ~ -3.3, log T(K) ~ 3.9, log P/k ~ -1.0 and [Si/H] = -0.14±0.23. The collisionally ionized gas with log N(O VI) =14.59±0.07 and log N(C IV) =14.14±0.05 likely arises in cooling transition temperature gas with log T ~ 5.5. The baryonic content of the warm gas exceeds that in the cool gas by a large factor estimated to range from 6 to 66.

2. The z = 0.06342 absorber is associated with a L ~ 0.31L* spiral emission line galaxy with an impact parameter of 63 kpc and an optical spectrum dominated by A stars. The absorber is tracing photoionized gas with solar elemental abundances and collisionally ionized gas of unknown abundances. The absorber likely traces gas in the halo of the spiral emission line galaxy that has been enriched by gas ejected from the emission line galaxy over the last several Gyrs. The observations imply modest luminosity spiral galaxies are capable of ejecting matter to large distances.

3. The simple absorption system at z = 0.14072 only contains O VI and broad and narrow H I. The narrow H I implies cool gas with log T < 4.67. The O VI absorption with log N(O VI) = 13.60±0.09 and b = 33.4±11.9 km s$^{-1}$ is probably associated with the broad H I λ1215 absorption with log N(H I) = 13.47±0.10 and b = 86.7±15.4 km s$^{-1}$. The difference in Doppler parameters implies the detection of warm gas with log T = 5.61(+0.16, -0.25), log N(H) = 19.54(+0.26, -0.44) and [O/H] = -1.48 (+0.46, -0.26).

4. The z = 0.14072 absorber is possibly associated with a 1.6L* absorption line galaxy with an impact parameter of 617 kpc which is found among 6 moderate luminosity galaxies within 3.4 Mpc of the 3C 263 line of sight. In this case the absorber could be tracing a warm highly extended halo. However, the WIYN galaxy survey only reaches L ~ 0.34L* at this redshift so we can't rule out an association with a low luminosity galaxy or with a WHIM filament connecting to the galaxies.

5. BLAs associated with O VI absorption tracing warm intervening gas with 5 < log T <6.1 have clearly been detected in the five absorption systems listed in Table 7. The oxygen abundance in these systems ranges from [O/H] = -1.7 to -0.3. The total baryonic content of these systems is large with log N ranging from 19.6 to 20.4.

Acknowledgements: We thank the many people involved with designing and building COS and determining its performance characteristics. We appreciate the assistance from Drs. Christy Tremonti and Yamnei Chen in determining the properties of Galaxy G. Dr Steve Penton kindly provided the wedge diagram shown in Fig. 9. The anonymous referee was helpful in identifying parts of the manuscript the needed to be more clearly presented. BDS and TSK acknowledge funding support from NASA through the COS GTO contract to the University of Wisconsin-Madison through NASA grants NNX08AC146 and NAS5-98043 to the University of Colorado at Boulder.

Table 1. Survey Completeness for the 3C 263 Sight Line

| η | 5' | 10' | 15' | 20' |
|---|---|---|---|---|
| limiting g magnitude | total number galaxies (measured number redshifts) | | | |
| 19.0 | 2(2) | 8(8) | 16(16) | 27(27) |
| 19.5 | 3(3) | 19(19) | 34(34) | 54(53) |
| 20.0 | 8(8) | 36(36) | 69(65) | 119(109) |

Table 2
Profile Fit Results for the z = 0.06342 System[a]

| ion   | $\lambda_r$   | z         | $v$[b] (km s$^{-1}$) | b (km s$^{-1}$) | log N (dex)  | Note |
|-------|---------------|-----------|----------------------|-----------------|--------------|------|
| H I   | 1215 to 972   | =0.063511 | = 27                 | 20.5±2.1        | 15.15±0.13   | 1    |
| "     | "             | =0.063390 | = -9                 | 19.3:±17.6      | 14.30:±0.36  | 1    |
| "     | "             | =0.063278 | = -43                | 44.9±7.6        | 15.04±0.10   | 1    |
| "     | "             | 0.062883  | -161±3±10            | 27.0±1.9        | 13.92±0.05   | 2    |
| "     | "             | 0.063761  | 102±4±10             | 12.7±9.8        | 12.70±0.14   | 2    |
| C II  | 1334          | 0.063499  | 23±2±10              | 17.1±2.9        | 13.37±0.04   |      |
| Si II | 1260, 1193    | 0.063482  | 18±3±10              | 11.8±5.9        | 12.26±0.09   |      |
| Si III| 1206          | 0.063507  | 26±1±10              | 12.5±1.2        | 12.90±0.03   | 5    |
| "     | "             | 0.063251  | -51±2±10             | 22.1±2.6        | 12.65±0.03   | 5    |
| C IV  | 1548, 1550    | 0.063511  | 27±1±10              | 15.1±1.7        | 13.86±0.03   | 1    |
| "     | "             | 0.063390  | -9±3±10              | 9.9±7.4         | 13.45±0.22   | 1    |
| "     | "             | 0.063278  | -43±8±10             | 21.9±7.8        | 13.57±0.19   | 1    |
| "     | "             | …         | …                    | …               | 14.14±0.05   | 3    |
| O VI  | 1031, 1037    | 0.063421  | = 0±4±10             | 38.7±6.3        | 14.59±0.07   | 4    |

[a] The Voigt profle fit code VPFIT and the wavelength dependent COS LSFs from Kriss (2011) were used to obtain the component fit results listed for the COS measurements in this table. The FUSE O VI observations were fitted with a LSF assumed to be a Gaussian with FWHM = 20 km s$^{-1}$. The reference redshift of 0.06342 for the absorption system is based on the single component fit to the O VI $\lambda\lambda$1031, 1037 absorption. Absorption line wavelengths and f-values were taken from Morton (2003). The profile fits are shown in Figure 1. [b]Component velocities are listed based on the reference redshift of 0.063421 for the O VI absorption. The first velocity error is the profile fit statistical error the second error is the COS velocity calibration error which is estimated to be ±10 km s$^{-1}$. Within a particular ionic absorber, such as C IV and Si III, the component to component relative velocity errors are ~ ±5 km s$^{-1}$.

Notes: 1. The principal H I three component structure was assumed to have the same velocities as the well defined C IV component structure. The three H I components are strongly saturated in the lines of H I $\lambda\lambda$1215, 1025. Therefore, the H I column densities in the three components are mostly derived from the weaker H I $\lambda$972 absorption which is not strongly saturated. 2. Two extra components at negative and positive velocity not seen in C IV are required to fit the H I absorption. The component at z = 0.06376 has very uncertain properties. Although fitted by a narrow feature, this component might instead represent the positive velocity extent of a BLA associated with the warm gas in the absorption system. 3. The total C IV column density summed over the three components is listed. 4. A single component fit to the broad O VI absorption is listed. The column density error may be larger if the actual O VI component structure is more complex. 5. Si III is detected in two absorption components.

Table 3. Profile Fit Results for the z = 0.14072 System[a]

| ion | $\lambda_r$ | z | $v^b$ (km s$^{-1}$) | $b^c$ (km s$^{-1}$) | log N (dex) |
|---|---|---|---|---|---|
| H I | 1215 to 1025 | 0.140745 | 7±1±10 | 27.9±1.0 | 14.51±0.03 |
| " | " | 0.140700 | -7±7±10 | 86.7±14.6±4.8 | 13.47±0.10 |
| O VI | 1031, 1037 | 0.140723 | 0±6±10 | 33.4±11.9 | 13.60±0.09 |

[a] The Voigt profle fit code VFIT and the wavelength dependent COS LSFs from Kriss (2011) were used to obtain the component fit results listed in this table. Absorption line wavelengths and f-values were taken from Morton (2003). The profile fits for O VI and H I are shown in Figure 3.

[b] Component velocities are listed based on the reference redshift of 0.14072 for the O VI absorption. The first velocity error is the profile fit statistical error the second error is the COS velocity calibration error which is estimated to be ±10 km s$^{-1}$. The narrow and broad H I absorption differ in velocity by 14±7 km s$^{-1}$ since the velocity calibration error should be the same for two closely spaced lines. However, the velocity calibration error must be included when intercomparing the H I and O VI absorption. Therefore v(O VI) – v(BLA) = 7 ± 17 km s$^{-1}$ while v(O VI) – v(narrow H I) = -7±16 km s$^{-1}$, assuming the different errors add in quadrature. Within the errors, the O VI could be either associated with the BLA or the narrow H I absorption.

[c] Profile fit errors are given. In the case of the BLA we list an additional continuum fit error of ± 4.8 km s$^{-1}$ derived by varying the wavelength position and widths of different continuum regions on each side of the absorption line located within the continuum windows shown in Figure 4.

Table 4. Galaxies within 500 km s$^{-1}$ of the z = 0.06342 Absorber

| RA (J2000.0) | Dec (J2000.0) | z | $\Delta v^a$ (km s$^{-1}$) | $\eta$ (arcmin) | $\rho h_{70}^{-1}$ (kpc) | g (mag) | $L^c$ (L$^*$) |
|---|---|---|---|---|---|---|---|
| 175.02154[b] | 65.80042[b] | 0.06322±0.00013 | -56±39 | 0.86 | 63 | 18.239±0.007 | 0.307 |
| 174.73408 | 65.87654 | 0.06281±0.00006 | -172±19 | 7.84 | 571 | 19.723±0.023 | 0.078 |
| 174.74136 | 65.88236 | 0.06285±0.00010 | -160±31 | 7.92 | 576 | 18.637±0.009 | 0.213 |
| 175.52030 | 65.81338 | 0.06498±0.00004 | 437±13 | 13.14 | 956 | 19.386±0.024 | 0.107 |
| 174.57166 | 65.58144 | 0.06248±0.00005 | -264±16 | 16.52 | 1203 | 19.385±0.014 | 0.107 |

[a] $\Delta v$ = v(galaxy) – v(absorber) in the rest fame of the absorber.
[b] The first galaxy listed in Galaxy G (see Section 5.1).
[c] The WIYN survey is 100% complete for $\eta$ < 10' to L ~ 0.05L$^*$ at z ~ 0.063.

Table 5. Galaxies within 500 km s$^{-1}$ of the z = 0.14072 Absorber

| RA (J2000.0) | Dec (J2000.0) | z | $\Delta v^a$ (km s$^{-1}$) | $\eta$ (arcmin) | $\rho h_{70}^{-1}$ (kpc) | g (mag) | $L^b$ (L$^*$) |
|---|---|---|---|---|---|---|---|
| 175.03847 | 65.73135 | 0.14087±0.00023 | 38±71 | 4.14 | 617 | 18.304±0.008 | 1.607 |
| 175.08533 | 65.66093 | 0.14095±0.00005 | 59±21 | 8.52 | 1269 | 19.230±0.016 | 0.685 |
| 175.02955 | 65.97460 | 0.14194±0.00010 | 314±34 | 10.70 | 1593 | 18.845±0.015 | 0.977 |
| 175.05491 | 65.99612 | 0.14169±0.00047 | 250±142 | 12.05 | 1795 | 19.836±0.022 | 0.392 |
| 175.77323 | 65.83538 | 0.14124±0.00006 | 134±23 | 19.45 | 2896 | 20.181±0.025 | 0.285 |
| 174.23793 | 65.57655 | 0.13998±0.00033 | -192±100 | 22.76 | 3390 | 19.327±0.480 | 0.627 |

[a] $\Delta v$ = v(galaxy) – v(absorber) in the rest fame of the absorber.
[b] The WIYN survey is 100% complete for $\eta$ < 10' to L ~ 0.34L$^*$ at z ~ 0.14.

Table 6. Emission Line Fluxes for Galaxy G at z = 0.06322

| Line | I ($10^{-16}$ erg cm$^{-2}$ s$^{-1}$) observed[a] | I ($10^{-16}$ erg cm$^{-2}$ s$^{-1}$) corrected for $\tau_V = 0.54$ | Note |
|---|---|---|---|
| [O II] $\lambda$3727 | 268±4: | 544±8: | 1 |
| H$\gamma$ | 36.9±1.4 | 69.7±2.7 | 2 |
| H$\beta$ | 82.5±1.5 | 148.6±2.7 | 2 |
| [O III] $\lambda$4959 | 20.3±1.1 | 36.3±2.0 | |
| [O III] $\lambda$5007 | 59.9±1.2 | 106.7±2.2 | |

[a] These observed emission line intensities have been corrected for a small amount of galactic foreground extinction with $A_V$ =0.037.

Notes: 1. Measurements below 4000 Å are uncertain because of degraded resolution, large radiometric calibration errors, and large wavelength calibration errors. 2. The observed Balmer line emission has been corrected for the underlying stellar absorption through a stellar population model fit to the stellar spectrum which is dominated by A stars.

Table 7. Low z Absorption Systems Containing Warm Plasma with T > $10^5$ K and Measures of [O/H][a]

| QSO | z | log N(OVI) | log N(H I)$_{BLA}$ | log N(H) | log T(K) | [O/H] | ($\rho h_{70}^{-1}$)[b] (kpc) | Note[c] |
|---|---|---|---|---|---|---|---|---|
| MRK 290 | 0.01027 | 13.80±0.05 | 14.35±0.01 | ~19.6 | ~5.15 | ~ -1.7 | 424 | 1 |
| HE 0153-4520 | 0.22601 | 14.21±0.02 | 13.70±0.08 | 20.41(+0.13, -0.17) | 6.07(+0.09, -0.13) | -0.28 (+0.09, -0.08) | ? | 2 |
| HE 0226-4110 | 0.20701 | 14.37±0.03 | 13.87±0.08 | 20.06±0.09 | 5.68±0.02 | -0.89±0.15 | 38, 109 | 3 |
| 3C 263 | 0.14072 | 13.60±0.09 | 13.47±0.10 | 19.54(+0.26, -0.44) | 5.61(+0.16, -0.25) | -1.48 (+0.46, -0.26) | 617 | 4 |
| PG 1444+407 | 0.22032 | 13.94±0.07 | 13.65±0.05 | 19.69(+0.25, -0.48) | 5.59(+0.15, -0.25) | -1.37(+0.60, -0.20) | ? | 5 |

[a] Low redshift O VI absorption systems for which a careful analysis provides strong evidence for the presence of gas with log T > 5 in which the abundance of oxygen can be measured. The O VI appears to be associated with a BLA in all cases.

[b] Impact parameter of the closest known galaxy to the line of sight.

[c] Notes: (1) Measurements are from Narayanan et al. 2010. At the low implied temperature, the non-equilibrium ionization of the gas must be considered to obtain log N(H) and [O/H]. The absorber may be tracing the gas associated with a luminous galaxy with $\rho$ = 424 kpc or the warm gas in a galaxy filament. (2) Measurements are from Savage et al. 2011a. Deep galaxy redshift observations have not been obtained to search for the associated galaxy. (3) Measurements are from Savage et al. 2011b. Mulchaey & Chen (2009) have identified the two associated galaxies with impact parameters of 38 and 109 kpc. The small temperature error is derived from the constraint provided by the O VI and Ne VIII observations assuming both ions are formed in the same gas. (4) Measurements are from this paper. The galaxy search is sensitive to L ~ 0.3L*. (5) The profile fit measurements for this system are given in Tripp et al. (2008, see their figure 41 and Table 3). The H I absorption is a BLA with v = 3±8 km s$^{-1}$, b = 86±15 km s$^{-1}$, and log N(H I) = 13.65±0.05. The O VI absorption is well aligned with v = 0±5 km s$^{-1}$, b = 36±8 km s$^{-1}$ and log N(O VI) = 13.94±0.07. The values of log N and [O/H] in the table were obtained using the analysis methods for the system at z = 0.14072 in this paper. The large implied temperature and low oxygen abundance obtained imply that CIE is a good approximation for estimating [O/H]. The non-thermal broadening is estimated to be $b_{NT}$ = 29.8±10.7. It would be valuable to observe this system with COS at higher S/N to confirm that the H I absorption is well described by a single broad component.

Figures

Figure 1a and 1b.  Absorption line profiles in the $z = 0.06342$ system for detected species.  Continuum normalized intensity is plotted against velocity with a heliocentric reference redshift of  0.06342.  The profiles for O VI $\lambda\lambda$1031,1037, H I Ly $\beta$, $\gamma$ are from FUSE.  The other profiles are from COS.  The FUSE and COS resolutions are ~20 km s$^{-1}$ and ~18 km s$^{-1}$.  The FUSE data are binned to ~10 km s$^{-1}$ while the higher S/N COS data are binned to ~2.3 km s$^{-1}$. The Voigt profile component fits to the absorption are shown with the fit results listed in Table 2.   The lowest panel in Fig. 1b shows the alternate results of the  fit to the  H I $\lambda$1215 absorption including a possible BLA centered on $v = 0$ km s$^{-1}$  with b =79.2±15.4 km s$^{-1}$ instead of a single narrow  H I component at 102 km s$^{-1}$ with b = 12.7±9.8 km s$^{-1}$.

Figure 2.  A simple single slab photoionization model for the moderately ionized gas in the $z =0.06342$ absorber for the component with log N(H I) = 15.15±0.13.  The radiation background is from the Haardt & Madau (2001) model interpolated to $z = 0.063$ and includes ionizing photons from AGNs and star forming galaxies.  log N(X) is plotted against the logarithm of the photoionization parameter log U.   Heavy solid lines on the lighter curves for the various ions display the observed column density range.  The observed column densities for Si II, Si III and C II are consistent with log U = -2.57±0.10 and [Si/H] ~ -0.14 and [C/H] ~ 0.05 with reference solar abundances from Aspland et al. (2009).

Figure 3. Absorption line profiles in the $z = 0.14072$ system for detected species observed by COS.   Continuum normalized intensity is plotted against velocity with a reference redshift of  0.14072. The Voigt profile component fits to the absorption are shown with the fit results listed in Table 3.

Figure 4. Flux versus heliocentric wavelength for the COS observations of 3C 263 from 1382 to 1392 Å containing the H I $\lambda$1215 $z = 0.14072$ absorption line near 1386.8 Å.  The two continuum windows and the profile fitting window are shown.

Figure 5.  SDSS r-band mosaic of the region surrounding 3C 263 with the positions of galaxies near the redshift of the QSO absorbers labeled. The image has a field of view of 40'x40' and is oriented north-up, east-left.  The quasar position is marked with a green star.  Galaxies at z~0.063 are marked with blue squares and galaxies at z~0.140 are marked with red circles.  All galaxies are labeled with their measured redshifts.  The large circle centered on the quasar corresponds to a radius of 200 kpc at a redshift of z=0.063 and the small circle corresponds to a radius of 200 kpc at a redshift of z=0.140.

Figure  6. The HST/WFPCS F675 image of the 3C 263 field.  The QSO and Galaxy G with an impact parameter of 63 kpc are identified.  The insert on the upper right shows an enlarged view of the structure in Galaxy G.

Figure 7. WIYN spectrum of the $L = 0.31L^*$ blue emission line Galaxy G with an impact parameter of 63 kpc that is likely associated with the multiphase O VI absorber at $z = 0.06342$. The higher Balmer absorption lines imply the presence of an A star stellar population implying vigorous star formation over the the recent several Gyrs. Lines marked with an X are the residuals of the sky subtraction.

Figure 8. WIYN spectrum of the $L = 1.6L^*$ absorption line galaxy with $z = 0.14087$ and an impact parameter of 617 kpc possibly associated with the O VI absorber at $z = 0.14072$. Various redshifted absorption lines are identified. The line marked with an X is the residual of the sky subtraction.

Figure 9. The pie chart shows a 2-degree wedge in RA and Dec centered on the 3C 263 sight line. The left plot shows the galaxy distribution as a function of RA (collapsed along the Dec dimension) and the right plot shows the distribution as a function of Dec (collapsed along the RA dimension). Redshift increases in the radial direction and arcs of constant recession velocity are shown for reference. Galaxies plotted with "c" symbols are from the CfA redshift survey, "s" symbols are from SDSS, and "k" symbols are from the galaxy survey work of this paper. Galaxies that are close to the sight line (impact parameter < 1 Mpc) are shown with magenta symbols. The blue circles mark the absorbers found at redshifts near $z = 0.063$ and $z = 0.141$. The WIYN survey for galaxies observed at these redshifts is 100% complete for $L \sim 0.05L^*$ and $0.34L^*$, respectively.

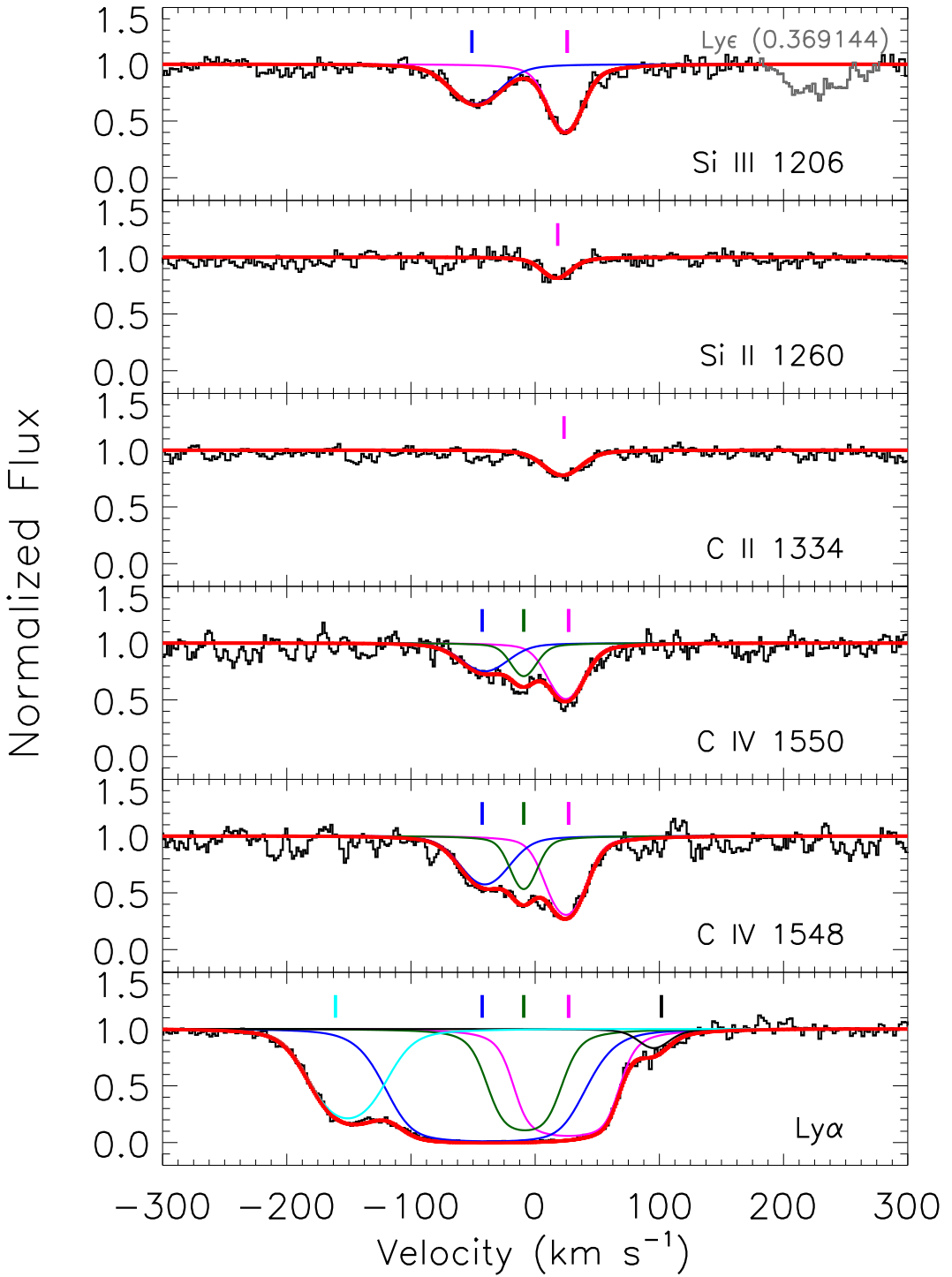

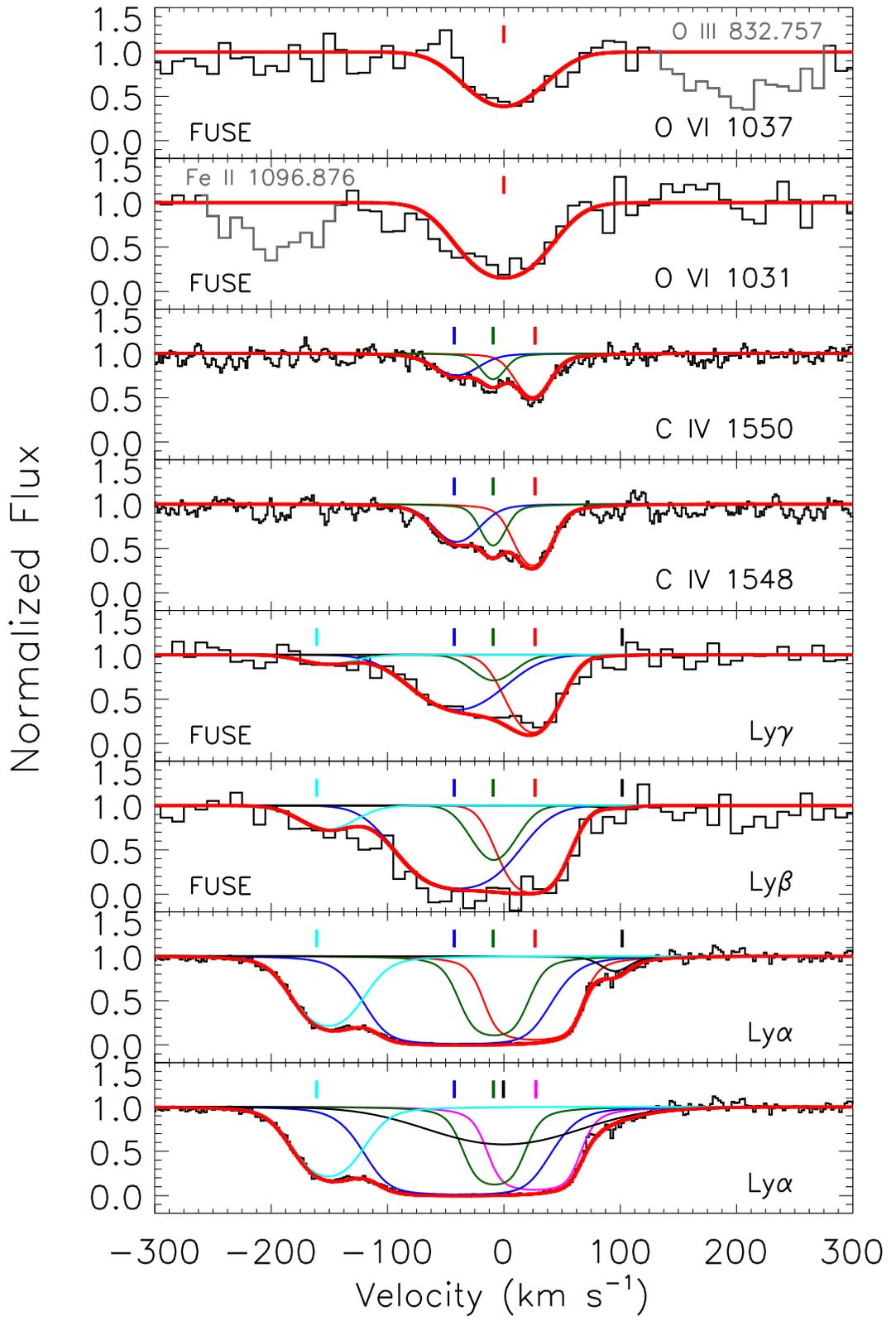

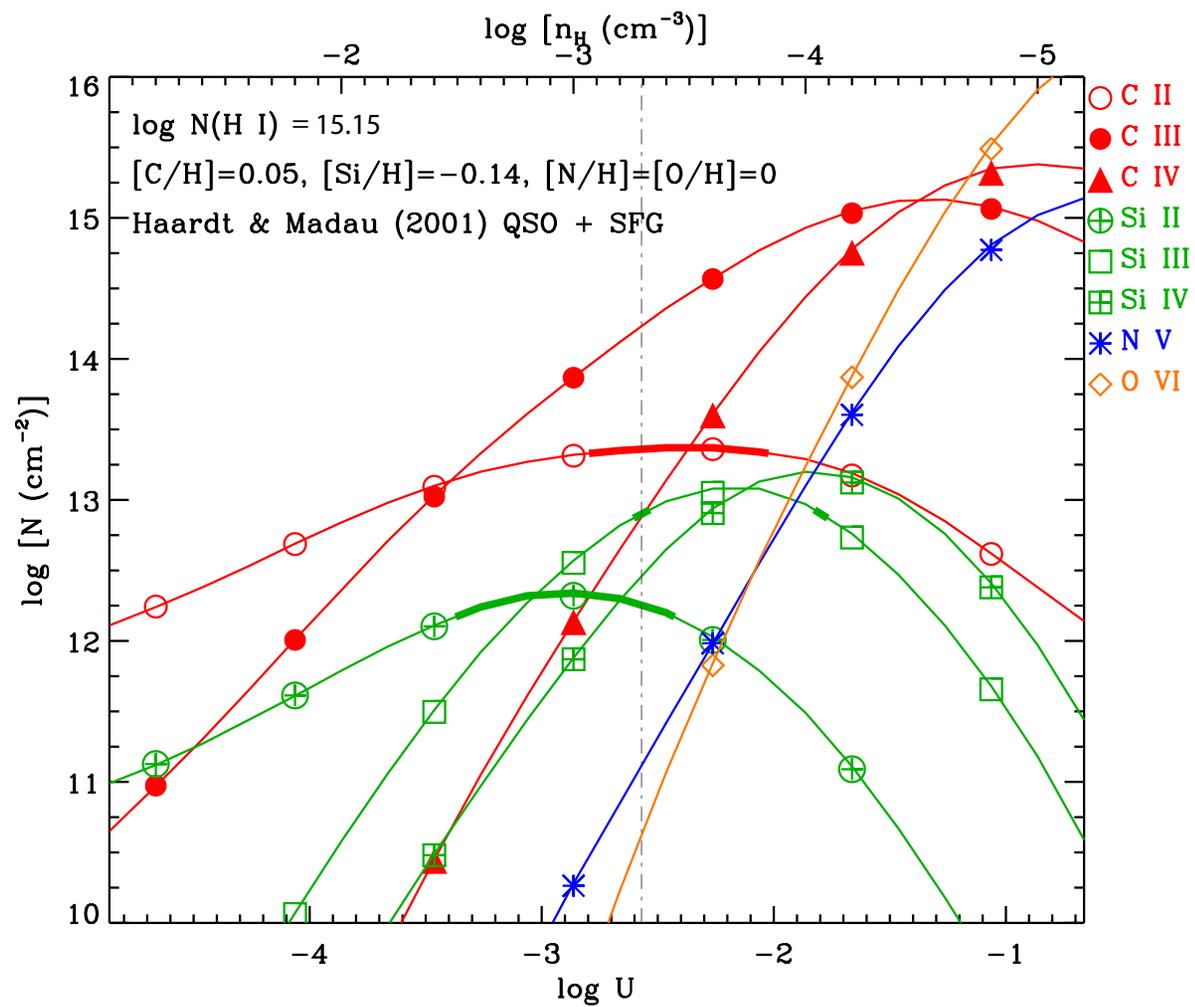

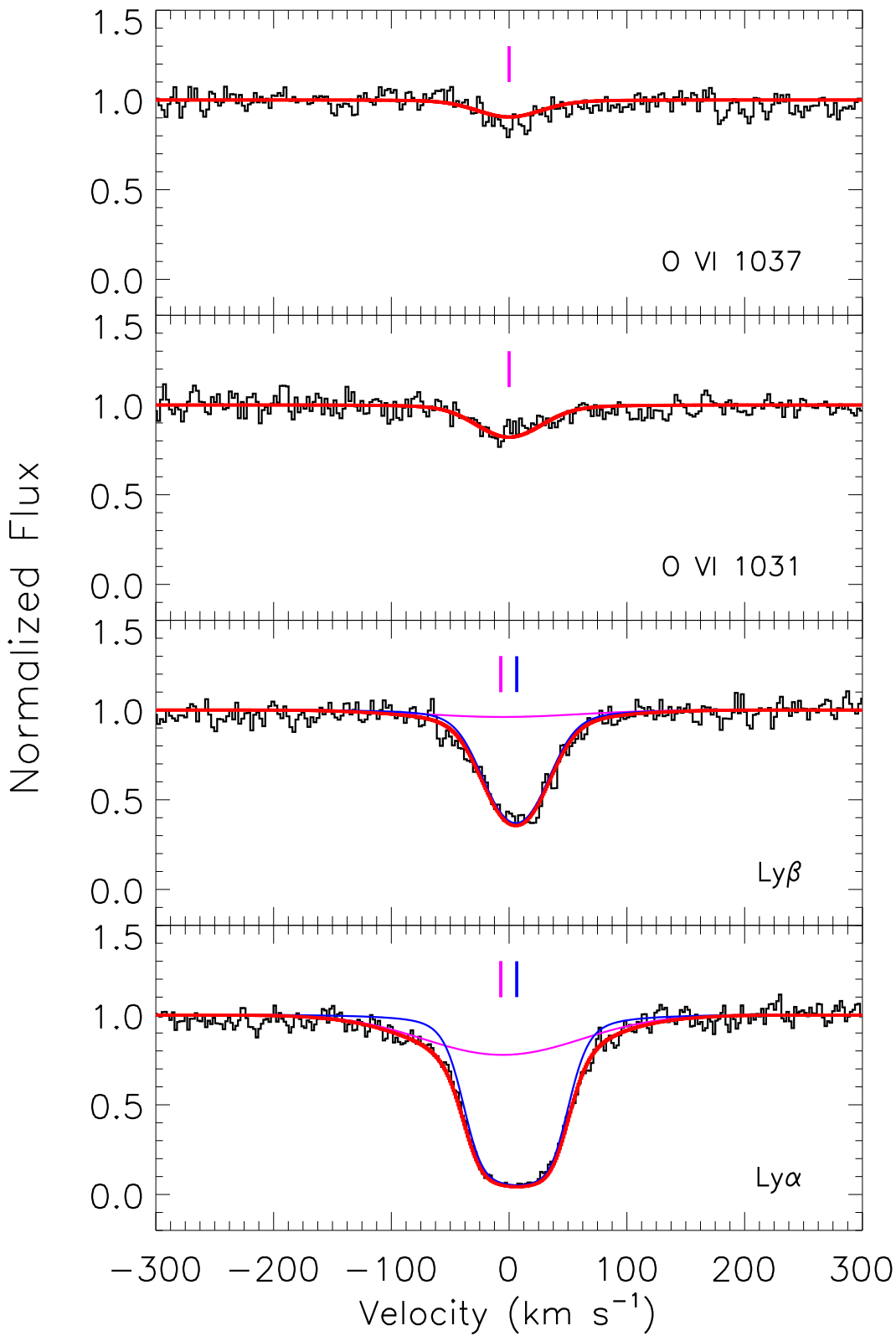

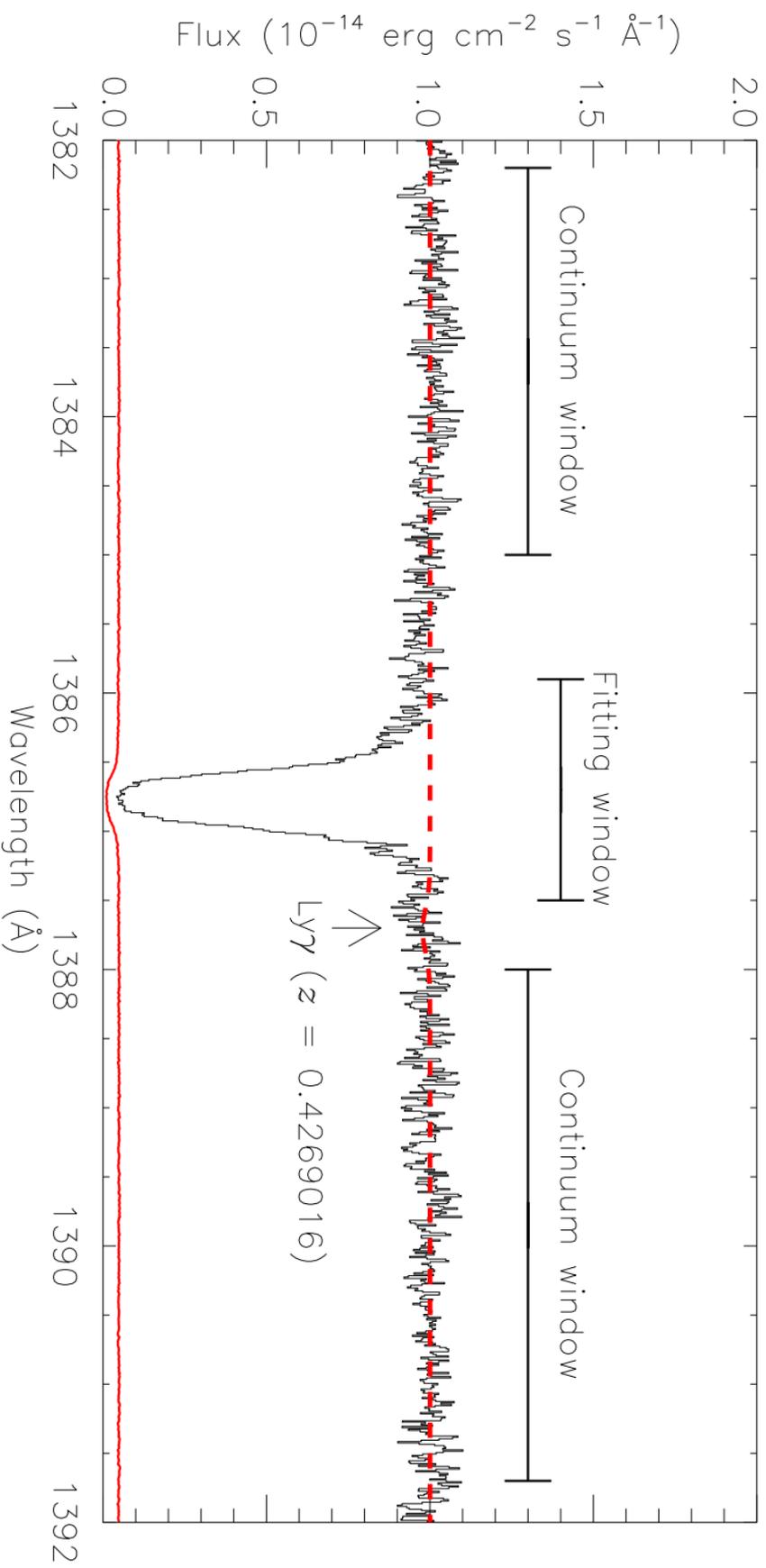

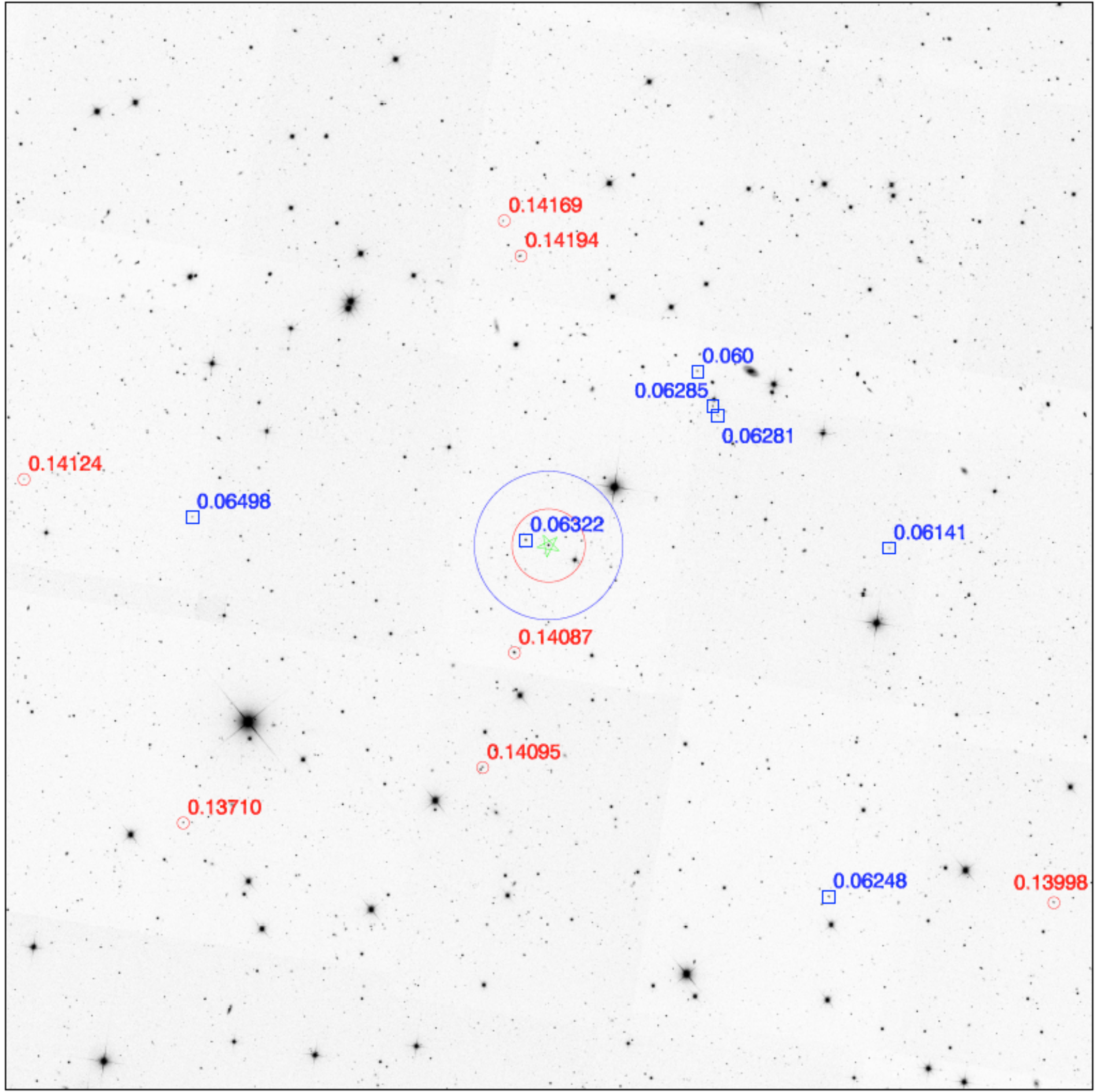

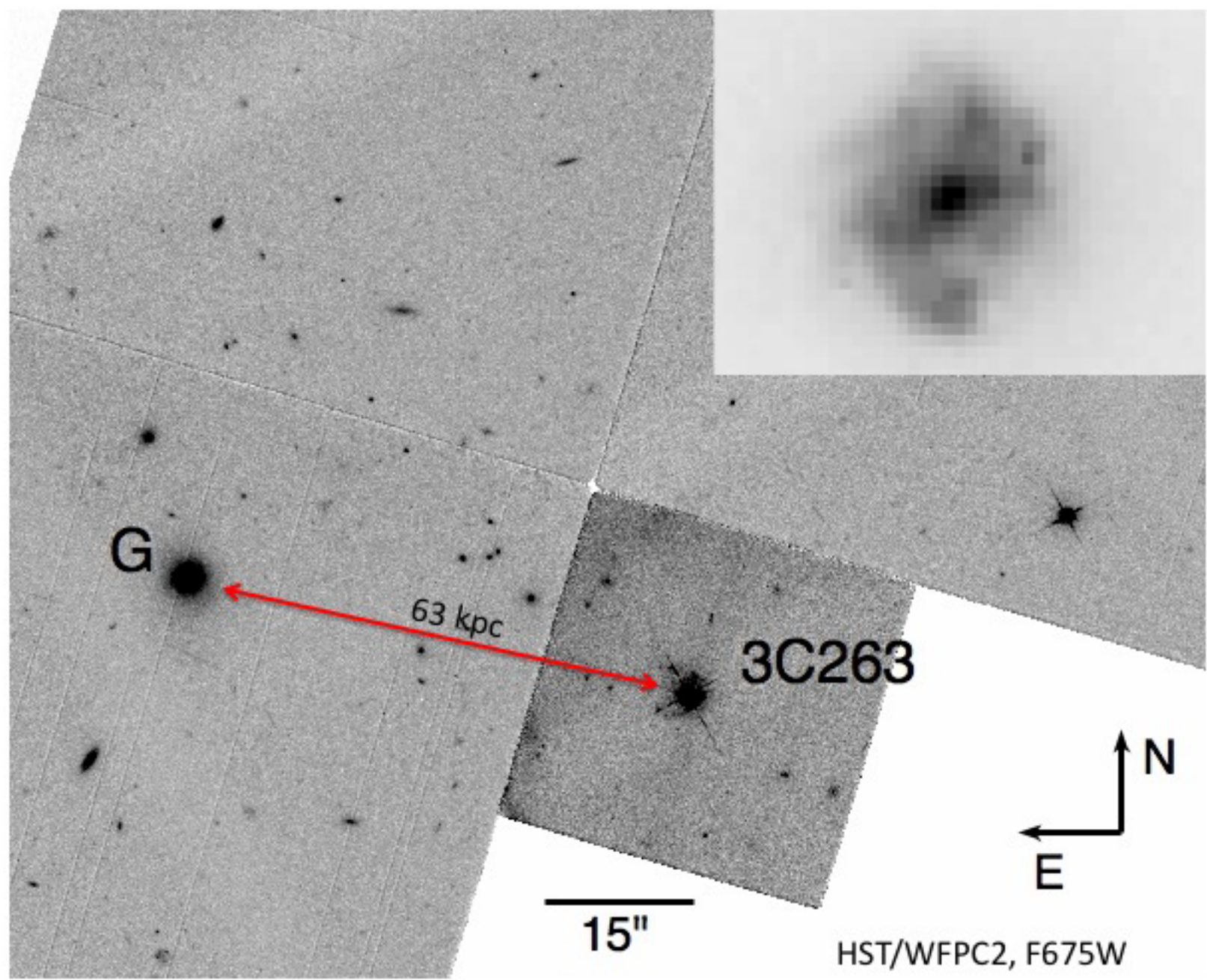

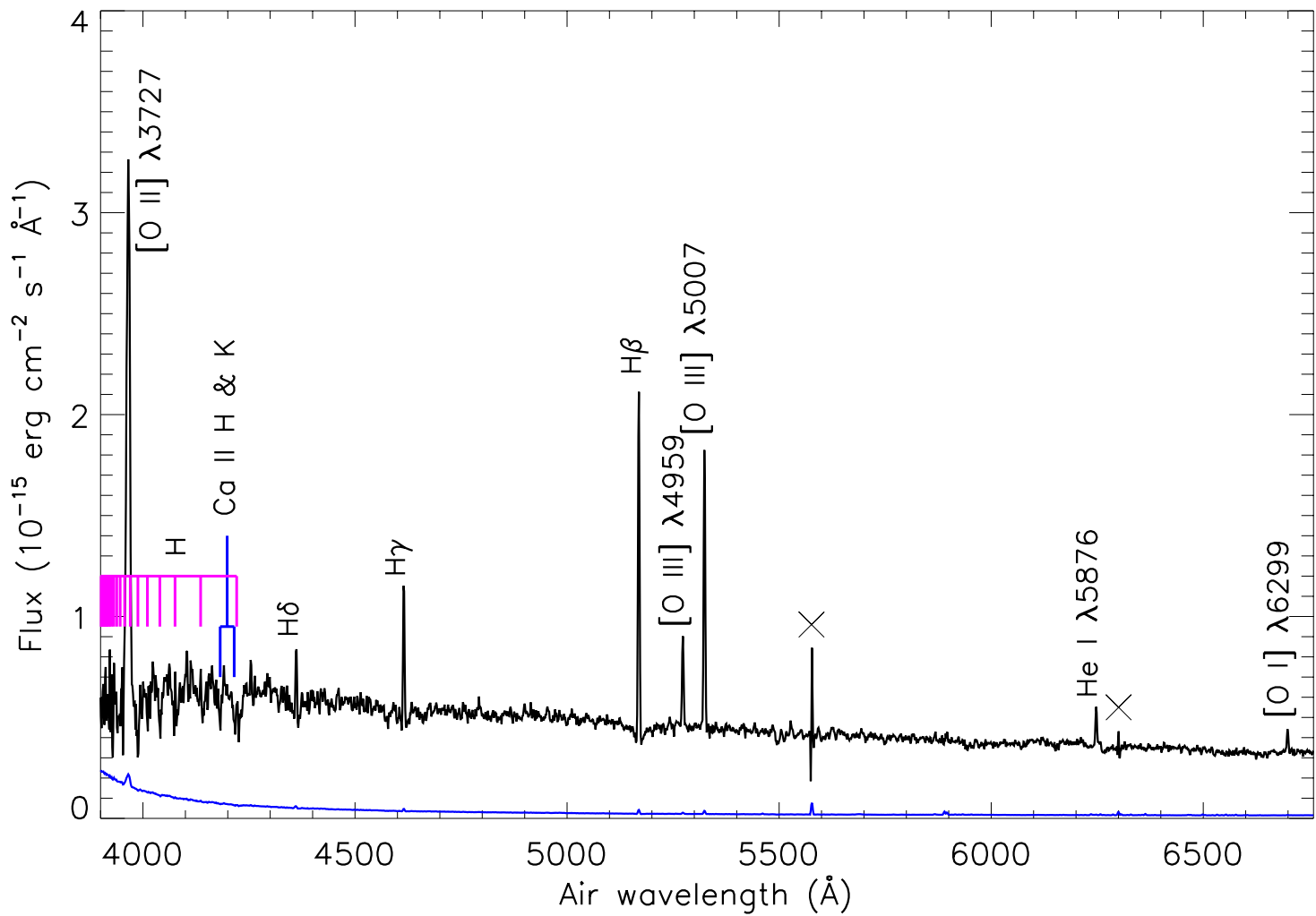

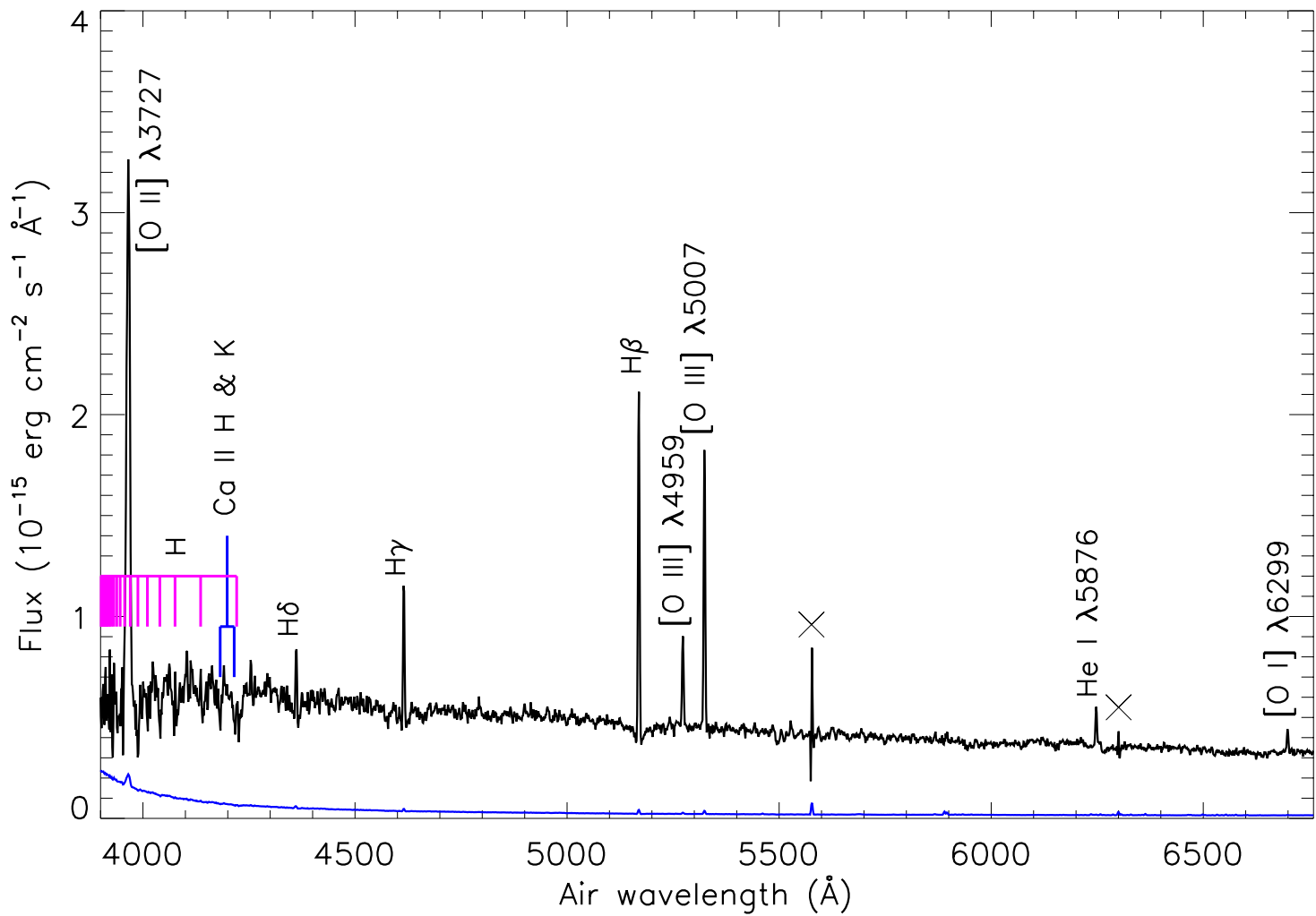

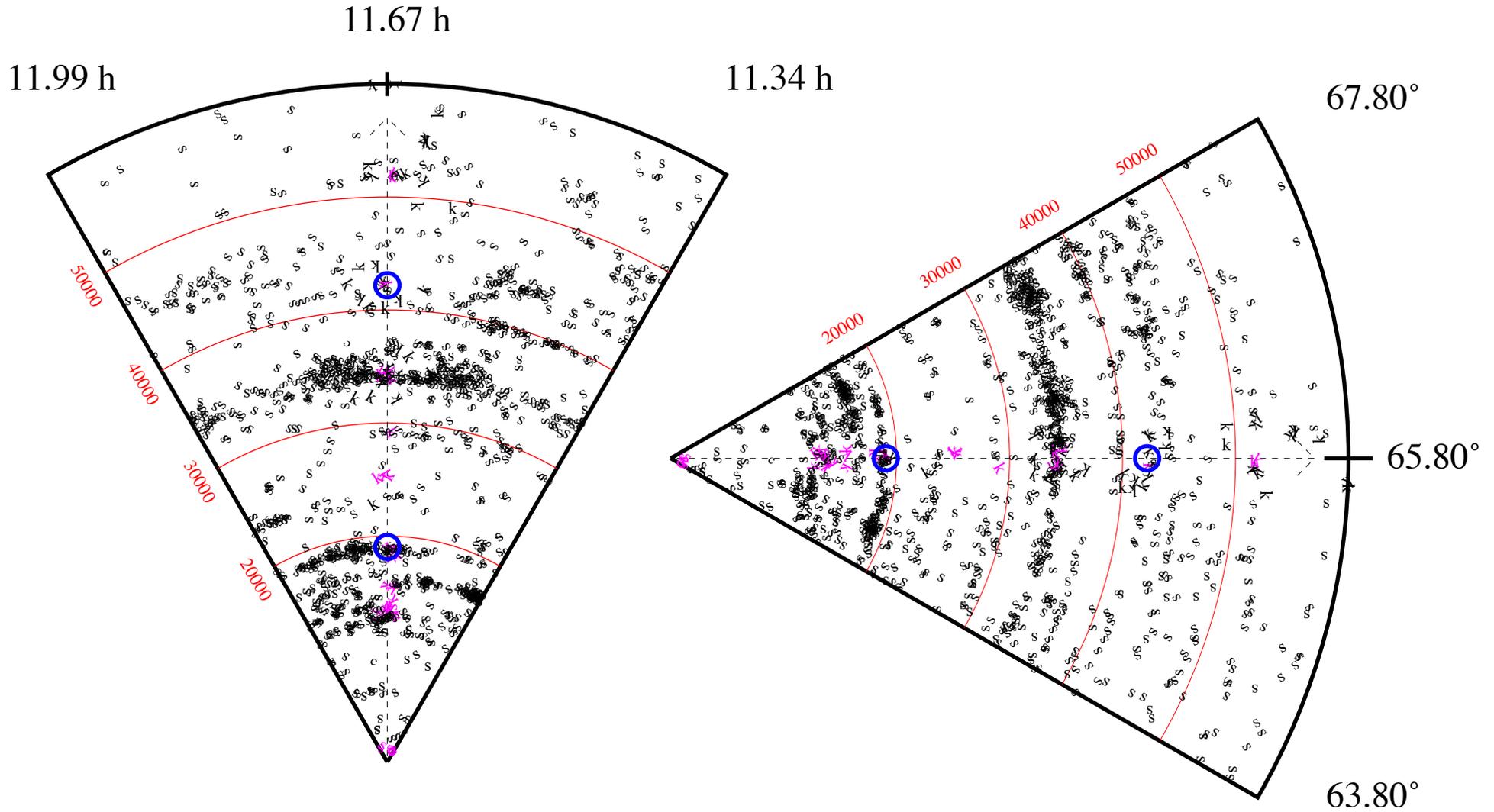